\documentclass[a4paper,aps,prd,showpacs,preprintnumbers,twocolumn,nofootinbib,superscriptaddress]{revtex4-1}

\usepackage[english]{babel}

\usepackage[utf8x]{inputenc}
\usepackage[T1]{fontenc}
\usepackage{ucs}
\usepackage{microtype}
\usepackage{newtxtext,newtxmath}
\usepackage{color}

\usepackage{url}
\usepackage{hyperref}

\usepackage{xspace}
\usepackage{lastpage}
\usepackage{stmaryrd}

\usepackage{graphicx}
\usepackage{empheq}
\usepackage{ifthen}
\usepackage{tensor}
\usepackage{paralist}

\usepackage{amsmath}
\usepackage{amsfonts}
\usepackage{amssymb}
\usepackage{mathrsfs}
\usepackage{extarrows}
\usepackage{verbatim}
\usepackage{upgreek}
\usepackage{wasysym}

\usepackage{natbib}

\newcommand\ph{\ensuremath{\varphi}}
\newcommand\eps{\ensuremath{\varepsilon}}

\newcommand{\cst}{\mathrm{cst}}

\newcommand\define{\equiv}

\newcommand\vect[1]{\boldsymbol{#1}}
\newcommand{\mat}[1]{\boldsymbol{#1}}
\newcommand\cplx[1]{\underline{#1}}
\newcommand\ex[1]{\mathrm{e}^{#1}}
\renewcommand\i{\ensuremath{\mathrm{i}}}

\renewcommand\Re{\ensuremath{\mathrm{Re}}}
\renewcommand\Im{\ensuremath{\mathrm{Im}}}

\newcommand\transpose[1]{#1^{\rm T}}
\newcommand{\tr}{\mathrm{tr}}

\newcommand\e[1]{_{\text{#1}}}
\newcommand\h[1]{^{\text{#1}}}
\newcommand\U[1]{\:\mathrm{#1}}

\newcommand{\dd}{\mathrm{d}}

\newcommand{\pd}[3][]{\frac{\partial^{#1} #2}{\partial {#3}^{#1}}}

\renewcommand\lim[2]{\underset{ #1 \rightarrow #2 }{ \mathrm{lim} } \,}

\newcommand{\delimiters}[4][]{
\ifthenelse{ \equal{#1}{1} }{  #2 #3 #4  }
					{ \ifthenelse{\equal{#1}{2}}{ \big#2 #3 \big#4 }
						{ \ifthenelse{\equal{#1}{3}}{ \Big#2 #3 \Big#4 }
							{ \ifthenelse{\equal{#1}{4}}{ \bigg#2 #3 \bigg#4 }
								{ \ifthenelse{\equal{#1}{5}}{ \Bigg#2 #3 \Bigg#4 }
									{ \left#2 #3 \right#4 }
								}
							}
						}
					}
													}

\newcommand{\pa}[2][]{\delimiters[#1]{(}{#2}{)}}
\newcommand{\pac}[2][]{\delimiters[#1]{[}{#2}{]}}

\newcommand{\abs}[2][]{\delimiters[#1]{|}{#2}{|}}
\newcommand{\ev}[2][]{\delimiters[#1]{\langle}{#2}{\rangle}}

\newcommand{\aap}{A{\&}A}

\newcommand{\mnras}{MNRAS}

\newcommand{\physrep}{Phys. Rep.}

\newcommand{\jacobi}{\mathcal{D}}
\newcommand{\amplification}{\mathcal{A}}
\newcommand{\source}{\mathcal{S}}
\newcommand{\image}{\mathcal{I}}
\newcommand{\beam}{\mathcal{B}}
\newcommand{\interior}{\mathrm{int}}
\newcommand{\exterior}{\mathrm{ext}}
\newcommand{\cir}{\mathscr{C}}
\newcommand{\disk}{\mathscr{D}}
\newcommand{\lens}{\lambda}
\newcommand{\quadrupole}{\mathcal{Q}}

\newcommand{\FLUletter}{FLU17\xspace}
\newcommand{\FLUadvanced}{FLU18b\xspace}

\begin{document}

\title{Cosmic convergence and shear with extended sources}

\author{Pierre Fleury}
\email{pierre.fleury@unige.ch}
\affiliation{D\'{e}partment de Physique Th\'{e}orique, Universit\'{e} de Gen\`{e}ve,\\
24 quai Ernest-Ansermet, 1211 Gen\`{e}ve 4, Switzerland}

\author{Julien Larena}
\email{julien.larena@uct.ac.za}
\affiliation{Department of Mathematics and Applied Mathematics\\
University of Cape Town,
Rondebosch 7701, South Africa}

\author{Jean-Philippe Uzan}
\email{uzan@iap.fr}
\affiliation{
            Institut d'Astrophysique de Paris, CNRS UMR 7095, Universit\'e Pierre et Marie Curie--Paris VI, 98 bis Boulevard Arago, 75014 Paris, France, and \\
           Sorbonne Universit\'es, Institut Lagrange de Paris, 98 bis, Boulevard Arago, 75014 Paris, France}

\begin{abstract}
The standard theory of weak gravitational lensing relies on the approximation that light beams are infinitesimal. Our recent work showed that the finite size of sources, and the associated light beams, can cause nonperturbative corrections to the weak-lensing convergence and shear. This article thoroughly investigates these corrections in a realistic cosmological model. The continuous transition from infinitesimal to finite beams is understood, and reveals that the previous results overestimated finite-beam effects due to simplistic assumptions on the distribution of matter in the Universe. In a KiloDegree Survey-like setting, finite-beam corrections to the cosmic shear remain subpercent, while percent-level corrections are only reached on subarcmin scales. This article thus demonstrates the validity of the infinitesimal-beam approximation in the interpretation of current weak-lensing data.
\end{abstract}

\date{\today}
\maketitle

\section{Introduction}

Weak gravitational lensing is one of the current key cosmological probes, together with the cosmic microwave background, type Ia supernovae, baryon acoustic oscillations, and other large-scale structure observables, such as redshift-space distortions. The main advantage of lensing resides in its sensitivity to all forms of energy, which allows one to map the distribution of matter in the Universe without relying on biased tracers, such as galaxies or neutral hydrogen. The near-past Canada-France-Hawaii Telescope Lensing Survey (CFHTLenS)~\cite{Fu:2014loa}, and the current KiloDegree Survey (KiDS)~\cite{Hildebrandt:2016iqg} and Dark Energy Survey (DES)~\cite{Abbott:2017wau}, have measured the combination of cosmological parameters~$\sigma_8\sqrt{\Omega\e{m}}$ with a precision of $3\%$; future surveys like Euclid~\cite{2010arXiv1001.0061R}, the Large Synoptic Survey Telescope~\cite{LSST}, or the Wide-Field InfraRed Survey Telescope~\cite{WFIRST}, are expected to improve those results by a factor of 10.

The theoretical framework of weak lensing is built upon the Sachs theory for the propagation of infinitesimal light beams~\cite{1961RSPSA.264..309S}, i.e., associated with infinitesimal sources. In that framework, light beams can only be focused and sheared by the local spacetime curvature that they experience. This leads to the two standard observables of weak lensing, namely convergence~$\kappa$ and shear~$\gamma$, which respectively characterize the magnification and elliptical deformation of images. This approximation is valid as long as the typical cross section of the light beams is much smaller than the scale over which their distortions vary appreciably. Extensions of the infinitesimal case have already been studied, including notably the arc-like deformations of images---the so-called flexion~\cite{Goldberg:2004hh, Bacon:2005qr}, related to the shear gradient. More generally, Refs.~\cite{2016CQGra..33pLT01C, 2016CQGra..33x5003C} developed an elaborate formalism to evaluate the normal modes of distortion of an image, based on series expansions of the curvature experienced by a beam.

However, because they are constructed from series expansions around the infinitesimal-beam case, the aforementioned approaches are unable to deal with light beams enclosing a distribution of matter with significant, unsmooth, variations of density. We proposed an alternative in Ref.~\cite{Fleury:2017owg}, hereafter \FLUletter, by adapting the strong-lensing formalism to weak lensing. This allowed us, as a by-product, to compute the finite-beam corrections to the weak-lensing shear. Quite unexpectedly, these corrections turned out to be large; in particular, we found in \FLUletter that in a Universe randomly filled with point lenses, the variance of the shear, $\ev[2]{|\gamma^2|}$ is no longer equal to the variance of the convergence, $\ev[2]{\kappa^2}$, as soon as the finiteness of the sources is taken into account. Instead, a factor $4/3$ appears: $\ev[2]{|\gamma|^2}=(4/3)\ev[2]{\kappa^2}$. This result suggested that there could be significant corrections to the standard interpretation of the weak-lensing data, thereby motivating a comprehensive investigation of weak lensing with extended sources, in order to determine the magnitude of finite-size effects accurately.

The present article reports the outcome of this investigation, where we carefully analyzed the finite-beam corrections to the weak-lensing convergence and shear in a cosmological model with realistic matter distribution. The motivation and content of our finite-beam formalism are exposed in Sec.~\ref{sec:finite-beam_formalism}, in much greater details than in \FLUletter. This formalism is applied to cosmology in Sec.~\ref{sec:cosmic_weak_lensing}, where we go from a discrete to a continuous distribution of lenses. The actual corrections to the standard weak-lensing observables, namely convergence and shear two-point correlations, are then computed in Sec.~\ref{sec:two-point_correlations}. In particular, we show how the $4/3$ factor found in \FLUletter was a consequence of the unrealistic Poisson distribution of matter that we adopted for simplicity in that first analysis. We finally evaluate the corrections to aperture mass measures in Sec.~\ref{sec:aperture_masses} before concluding in Sec.~\ref{sec:conclusion}.

Albeit theoretical, this article contains results which are relevant to current and future weak-lensing surveys. The slightly more technical companion article~\cite{advanced}, hereafter \FLUadvanced, encompasses the present analysis in a wider framework, thereby demonstrating the full potential of our formalism.

We adopt units in which $c=1$. Two-dimensional vectors are denoted with bold symbols ($\vect{\beta}, \vect{\theta}, \vect{\lambda}, \ldots$) while underlined quantities ($\cplx{\beta}, \cplx{\theta}, \cplx{\lambda}, \ldots$) are complex numbers canonically associated to them: if $\vect{\beta}=(\beta_x, \beta_y)$, then $\cplx{\beta}\define\beta_x + \i \beta_y$.

\section{Weak lensing with finite sources}
\label{sec:finite-beam_formalism}


\subsection{Scales, approximations, and regimes of lensing}

\begin{figure}[h!]
\centering
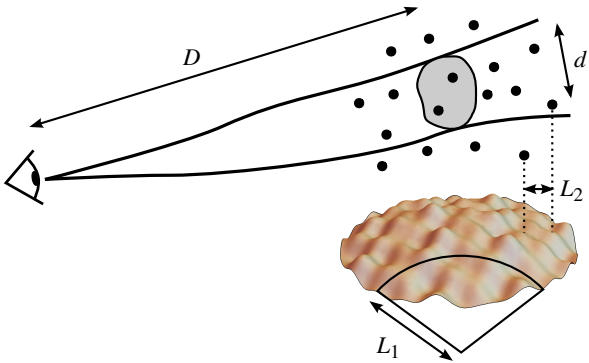
\caption{Fundamental length scales involved in geometric optics in curved spacetime.}
\label{fig:scales}
\end{figure}

Geometric optics in curved spacetime is usually characterized by a hierarchy of length scales. These are\footnote{The light's wavelength is not included in the list, because in the geometric optics regime (eikonal approximation) the wave nature of light is irrelevant.} (see Fig.~\ref{fig:scales}): the physical cross-sectional diameter~$d$ of the beam; the curvature radius~$D$ of the wave front, which is also the angular-diameter distance to the emergence or convergence point of the wave; and two spacetime quantities, namely its typical curvature radius~$L_1\sim (R_{\mu\nu\rho\sigma}R^{\mu\nu\rho\sigma})^{-1/4}$ and the scale over which it changes appreciably~$L_2\sim L_1/\partial L_1$. In general relativity, $L_1$ can be understood as a measure of the neighboring energy density~$\rho$ via $L_1\sim 1/\sqrt{G\rho}$, where $G$ denotes Newton's constant.\footnote{Strictly speaking, the local density of energy-momentum only sets the value of the Ricci part of the full Riemann curvature tensor, which excludes, in particular, the contribution of long-range tidal forces in $L_1$. However, the correspondence between $L_1$ and $\rho$ can be extended if $\rho$ is understood as the average energy density in a region containing the closest massive objects. For example, at a distance $r$ from a spherical mass $M$, $L_1\sim \sqrt{r^3/GM}$, where we can identify~$\rho\sim M/r^3$.} As a consequence, $L_2$ is related to the typical evolution scale of the energy density. The comparison of those scales allows one to make various approximations, as discussed below.

\subsubsection{Paraxial optics}

Paraxial optics---also known as optics in Gauss's conditions---assume that the beam's angular aperture is small, so that its cross section covers only a small fraction of the wave front, and can be considered flat. In other words,
\begin{equation}
d \ll D \ .
\end{equation}
In cosmology, this is known as the flat-sky approximation. Paraxial optics allows one to define a notion of optical axis, or line of sight, e.g. the central ray of the beam, which serves as a reference for the other rays. Perpendicularly to this common axis can be inserted a screen, on which the notions of size and shape of the light beam can be defined.

\subsubsection{Weak gravitational field}

Just like any object of size $d$, a light beam with diameter $d$ is said to experience a weak gravitational field if the associated tidal forces are small, that is
\begin{equation}
d \ll L_1 \ .
\end{equation}
Geometrically, this is equivalent to saying that, e.g., all events in the intersection between the light beam and a screen orthogonal to its axis (detection events) essentially belong to the same tangent hyperplane of the spacetime manifold. Note also that this assumption is necessary for the very notion of size~$d$ of the beam to be defined univocally.

\subsubsection{Locally homogeneous curvature}

Once virtual screens are placed all along the light beam, where its morphology is univocally defined, one would like to determine its evolution as light propagates, i.e. from a virtual screen to the next one. In the general-relativistic description of light beams, this is given by the geodesic deviation equation,
\begin{equation}\label{eq:gde}
k^\mu k^\nu \nabla_\mu \nabla_\nu \xi^\rho = - R\indices{^\rho_\mu_\sigma_\nu} k^\mu k^\nu \xi^\sigma \ ,
\end{equation}
where $k^\mu$ is the wave four-vector of a fiducial ray, and $\xi^\mu$ connects any ray of the beam to this fiducial ray. Clearly, Eq.~\eqref{eq:gde} can only be applied if the Riemann tensor is homogeneous across the beam, that is
\begin{equation}
d \ll L_2 \ .
\end{equation}

A beam can be treated as infinitesimal if the last three conditions are satisfied, $d\ll D, L_1, L_2$, but the latter is by far the most restrictive, and the least likely to hold in reality. Indeed, as soon as light propagates through matter, there is always some substructure on scales smaller than the beam itself. Going beyond the approximation of homogeneous curvature will thus be the focus of this article.

\subsubsection{Locally homogeneous Jacobi matrix}

Although the above assumption was formulated in general-relativistic terms, it can be rephrased in a more lensing-oriented way. Consider an extended source, where each point~$\vect{x}$ is observed in a direction\footnote{If the source has multiple images, the reasoning applies individually to each image.}~$\vect{\theta}(\vect{x})$. The Jacobi matrix of the map~$\vect{\theta} \mapsto \vect{x}(\vect{\theta})$ is defined as
\begin{equation}
\mat{\jacobi}(\vect{\theta}) = \pd{\vect{x}}{\vect{\theta}} \ .
\end{equation}
By definition, $\mat{\jacobi}$ relates the morphology of an infinitesimal image patch, $\dd^2\vect{\theta}$, observed in the direction~$\vect{\theta}$, to the morphology of the corresponding source patch, $\dd^2\vect{x}$.

Assuming that spacetime curvature is homogeneous across the beam is equivalent to stating that the Jacobi matrix~$\vect{\jacobi}$ is homogeneous across the image, that is, if $\Omega$ denotes the angular size of the image,
\begin{equation}
\Omega \ll |\vect{\jacobi}^{-1}\partial \vect{\jacobi} |^{-2} \ .
\end{equation}
Therefore, in the remainder of this article, we will make no distinction between the expressions finite light beam and extended source or image.

\subsubsection{Weak lensing}

Let us close this subsection with a word on the definition of weak lensing. Contrary to the notions of paraxial optics, weak field, or homogeneous curvature, weak lensing is a concept that requires the introduction of a background, i.e., a fiducial no-lensing situation with respect to which the strength of lensing can be evaluated. For instance, when dealing with microlensing in the Milky Way, the background is taken to be the Minkowski spacetime, while in cosmology one would choose the Friedmann-Lema\^itre-Robertson-Walker (FLRW) spacetime. Once a background is chosen, lensing is defined as the map between the direction(s)~$\vect{\theta}$ in which a point image is observed, and the direction~$\vect{\beta}$ in which it would be observed through the background spacetime. Lensing is then said to be weak if $\vect{\theta}\approx\vect{\beta}$, or, equivalently, if the amplification matrix
\begin{equation}
\mat{\amplification} = \pd{\vect{\beta}}{\vect{\theta}} = \bar{D}\e{A}^{-1} \mat{\jacobi} \ ,
\end{equation}
where $\bar{D}\e{A}$ is the background angular-diameter distance to the source, is close to unity all across the image.

\begin{figure}[h!]
\centering
\includegraphics[width=\columnwidth]{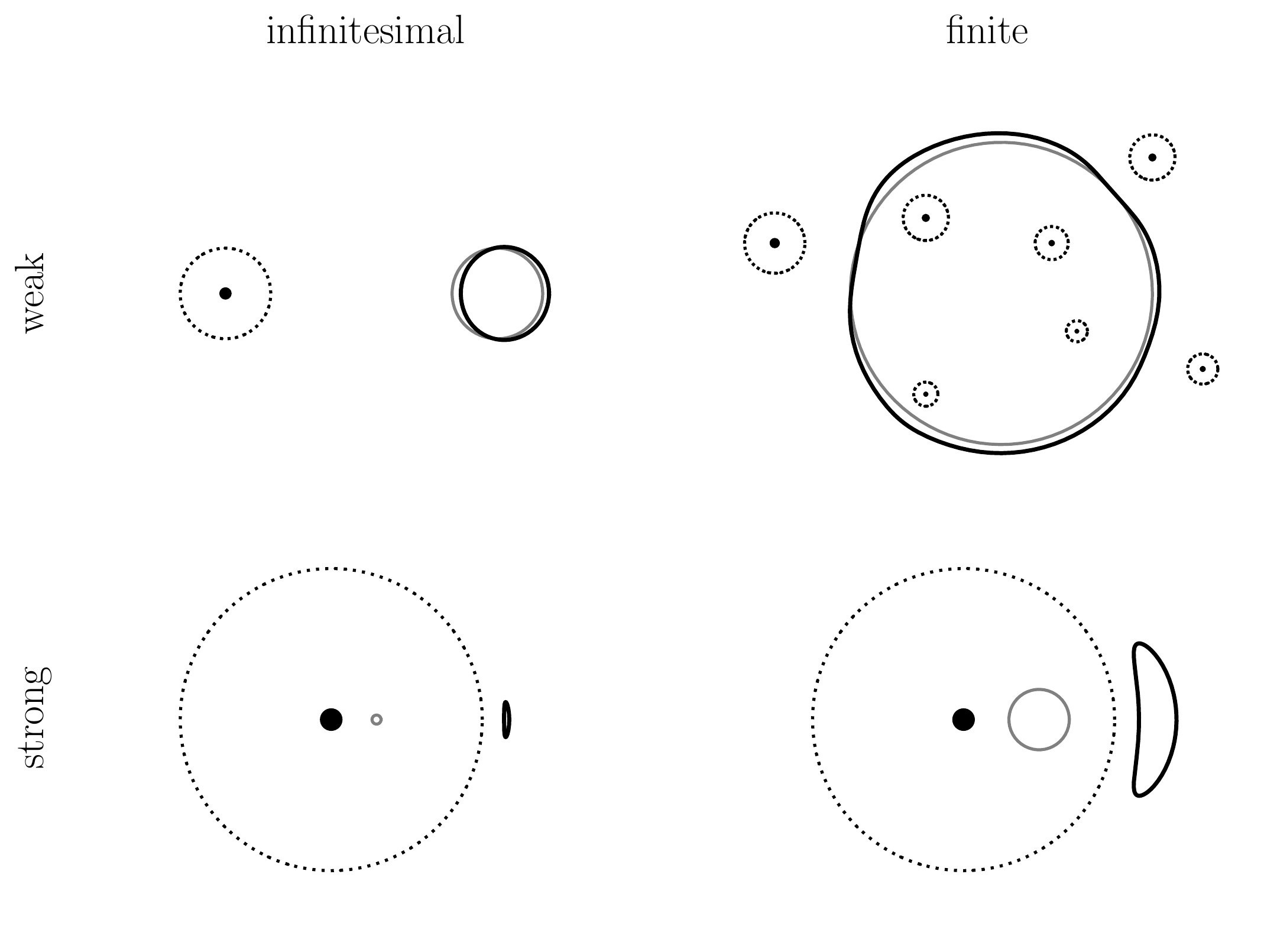}
\caption{Various regimes of gravitational lensing: weak/strong with infinitesimal/extended sources. In each figure, a gray solid line represents the contour of a circular source, a black solid line is the image by lenses shown by black dots, while their Einstein radii are depicted with dotted lines.}
\label{fig:lensing_regimes}
\end{figure}

The dichotomy between weak and strong lensing is thus quite different from the distinction between infinitesimal and finite sources, for the latter depends on the size of the source/beam, while the former does not. One can thereby envisage situations where strong-lensing effects are present without finite beams, and conversely, as illustrated in Fig.~\ref{fig:lensing_regimes}. Yet, confusingly enough, in the canonical example of the Schwarzschild lens, the transition between weak and strong lensing seems to coincide with the transition between infinitesimal and finite source. This is actually due to the power-law behavior of the gravitational potential generated by an isolated massive body: the shorter $L_1$, the shorter $L_2$; hence, the larger~$\mat{\amplification}$, the shorter the scale over which it varies.

The distinction between weak and strong lensings must not be confused either with the one between weak and strong gravitational field regimes. In practice, strong lensing events are always observed in situations where light beams traveled though weak gravitational fields only. This is because lensing is the cumulative effect of tidal forces acting on the light beam from the source to the observer.

In the remainder of this article, we will focus on weak lensing, in the paraxial and weak-field regimes, but with no assumption regarding the size of the light beam.

\subsection{Lens equation}

Consider a statistically homogeneous and isotropic Universe, made of noncompact, spherical, nonrotating, and slowly moving massive objects (apart from their cosmic recession). The associated spacetime geometry can then be described by the FLRW metric with scalar perturbations,
\begin{multline}
\dd s^2 = a^2(\eta) \Big\{ -(1+2\Phi)\dd\eta^2 \\
											+ (1-2\Phi) \pac{\dd\chi^2 + f_K^2(\chi) \, \dd\Omega^2 } \Big\} \ ,
\end{multline}
where $a$ denotes the scale factor quantifying cosmic expansion, $K$ is the background spatial curvature parameter, $f_K(\chi) \define \sin(\sqrt{K}\chi)/\sqrt{K}$, $\Phi$ is the Newtonian gravitational potential generated by the massive objects, and $\eta, \chi$ are respectively the background conformal time and comoving radial coordinate.

\begin{figure}[h!]
\centering
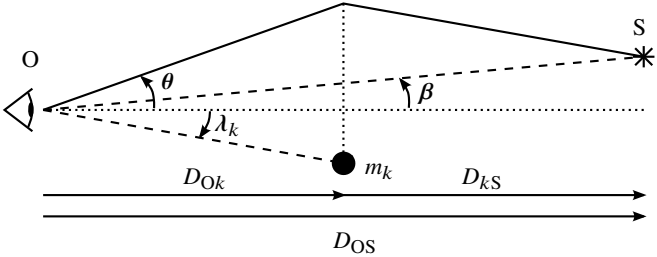
\caption{Geometric quantities involved in the lens equation.}
\label{fig:lens_equation}
\end{figure}

Solving the null geodesic equation in this spacetime then yields the relationship between the (lensed) direction~$\vect{\theta}$ in which an image is actually observed, with the (unlensed) direction~$\vect{\beta}$ in which it would be observed if all the matter forming the massive objects were homogeneously distributed in the Universe ($\Phi=0$). In paraxial optics, this relation reads~\cite{1992grle.book.....S}
\begin{equation}\label{eq:lens_equation}
\vect{\beta} = \vect{\theta} 
- \sum_k \eps_k^2 \, \frac{\vect{\theta}-\vect{\lens}_k}{\abs{\vect\theta-\vect{\lens}_k}^2} \ ,
\end{equation}
and is known as the lens equation (see Fig.~\ref{fig:lens_equation}). It involves the unlensed position~$\vect{\lens}_k$ of each lens~$k$ in the Universe, and its Einstein radius
\begin{equation}\label{eq:Einstein_radius}
\eps_k^2 \define \frac{4G m_k D_{k\mathrm{S}}}{D_{\mathrm{O}k} D\e{OS}}
= \frac{4 G m_k (1+z_k) f_K(\chi\e{S}-\chi_k)}{f_K(\chi\e{S}) f_K(\chi_k)} \ ,
\end{equation}
where $m_k$ is the mass of the lens, while $D_{k\mathrm{S}}$, $D_{\mathrm{O}k}$, and $D\e{OS}$ are the angular-diameter distances, respectively, of the source seen from the lens $k$, of the lens~$k$ seen from the observer, and of the source seen from the observer. Specifying the points of view is important here, because those distances are affected differently by aberration effects. This is why $z_k$ is in principle the true observed redshift of the lens~$k$, and not the background cosmological redshift associated to its radial position~$\chi_k$.

Since we have chosen the background to be FLRW, that is a Universe homogeneously filled with an energy density, the mass of the lenses~$m_k$, and hence the associated squared Einstein radius, $\eps_k^2$, are allowed to be \emph{negative}. This is due to the fact that the gravitational potential~$\Phi$ driving light deflection with respect to this background satisfies the Poisson equation in comoving coordinates~$\Delta\Phi = 4\pi G a^{2}(\rho-\bar{\rho})$, where $\bar{\rho}$ is the mean energy density. Introducing negative masses is an artificial trick to account for the presence of $-\bar{\rho}$, which will be of particular interest in Sec.~\ref{sec:discrete_to_continuous}; we refer the reader to Appendix~\ref{app:negative_lenses} for a more rigorous discussion.

Since we work with small angles, $\vect{\beta}, \vect{\theta}, \vect{\lens}_k$ can be considered small vectors on a plane orthogonal to the line of sight, spanned by an orthonormal basis $(\vect{e}_x, \vect{e}_y)$, to which we can associate complex numbers with the convention
\begin{equation}
\vect{\theta} = \theta_x \vect{e}_x + \theta_y \vect{e}_y
\longmapsto
\cplx{\theta} = \theta_x + \i \theta_y \ .
\end{equation}
The lens equation then becomes
\begin{equation}\label{eq:lens_equation_complex}
\cplx{\beta} = \cplx{\theta} - \sum_k \frac{\eps_k^2}{\cplx{\theta}^* - \cplx{\lens}_k^*} \ ,
\end{equation}
where a star denotes complex conjugation. This complex representation of gravitational lensing seems to have been first introduced in Ref.~\cite{1973ApJ...185..747B}.

In this article, we restrict to the weak-lensing regime, that is
\begin{equation}
|\mat{\amplification}| - 1 \sim \frac{\eps_k^2}{|\vect{\theta}-\vect{\lens}_k|^2} \ll 1 \ .
\end{equation}
In other words, the light rays of interest will always be far from the lenses' Einstein radii. In practice, the $\eps_k$ will be treated as very small numbers, with respect to which we can perform Taylor expansions.

\subsection{Convergence}

\begin{figure}[h!]
\centering
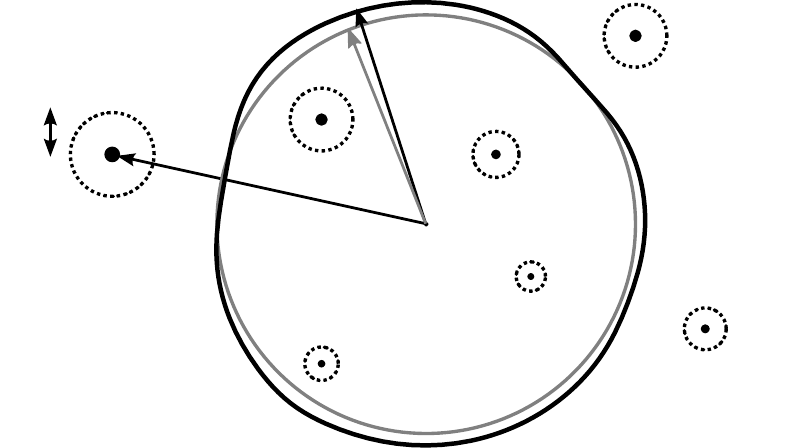
\caption{Image~$\image$ of an extended source~$\source$.}
\label{fig:extended_source}
\end{figure}

Let $\source$ be the contour of an extended source\footnote{By ``source,'' we actually mean the unlensed image.}, and $\image$ the contour of its image by the multiple lenses present on the way (see Fig.~\ref{fig:extended_source}). As mentioned above, for this situation to be in the weak-lensing regime, the image points~$\vect{\theta}$ must be only slightly displaced with respect to their unlensed counterpart~$\vect{\beta}$, which is satisfied if $\forall\vect{\beta}\in\source \ \forall k \ |\vect{\beta}-\vect{\lens}_k|\gg \eps_k$.

The weak-lensing convergence is defined by comparing the angular area~$\Omega$ of the image to the unlensed one~$\Omega\e{S}$,
\begin{equation}
\kappa \define \frac{\Omega - \Omega\e{S}}{2\Omega\e{S}} \ ,
\end{equation}
where $\Omega$ can be computed as
\begin{equation}\label{eq:Omega_def}
\Omega
= \int_{\interior\image} \dd^2 \vect{\theta}
= \frac{1}{2\i} \ointctrclockwise_{\mathcal{I}} \cplx{\theta}^* \dd \cplx{\theta} \ ,
\end{equation}
and similarly for $\Omega\e{S}$. Here, $\interior\image$ denotes the closed set of points in the complex plane located inside the contour $\image$. Using the lens equation~\eqref{eq:lens_equation_complex} to rewrite $\cplx{\theta}$ and $\dd\cplx{\theta}$ and in the expression \eqref{eq:Omega_def} of $\Omega$, yields the lowest order correction
\begin{multline}\label{eq:Omega_calc}
\Omega = 
\frac{1}{2\i} \ointctrclockwise_{\source} \cplx{\beta}^* \, \dd \cplx{\beta}
+ \frac{1}{2\i} \sum_k \eps_k^2 \ointctrclockwise_{\image} \frac{\dd \cplx{\theta}}{\cplx{\theta}-\cplx{\lens}_k}\\
- \frac{1}{2\i} \sum_k \eps_k^2 
	\pac{ \ointctrclockwise_{\image} \frac{\cplx{\theta}\dd \cplx{\theta}}{(\cplx{\theta}-\cplx{\lens}_k)^2} }^*
+ \mathcal{O}(\eps^4) \ .
\end{multline}
The first term on the right-hand side is $\Omega\e{S}$; the second term is readily evaluated using the residue theorem
\begin{equation}
\ointctrclockwise_{\image} \frac{\dd \cplx{\theta}}{\cplx{\theta}-\cplx{\lens}_k}
=
\begin{cases}
2\i\pi & \text{if $\cplx{\lens}_k$ is inside $\image$}\\
0 & \text{otherwise;}
\end{cases}
\end{equation}
as for the third term, since
\begin{equation}
\frac{\cplx{\theta}}{(\cplx{\theta}-\cplx{\lens}_k)^2}
= \frac{\cplx{\lens}_k}{(\cplx{\theta}-\cplx{\lens}_k)^2}
	+ \frac{1}{\cplx{\theta}-\cplx{\lens}_k} \ ,
\end{equation}
its residues are the same as the second term of Eq.~\eqref{eq:Omega_calc}, and we conclude that
\begin{equation}
\Omega = \Omega\e{S} + \sum_{k\in\interior\image} 2\pi \eps _k^2
+ \mathcal{O}(\eps^4) \ ,
\end{equation}
where $k\in\interior\image$ means that the lens $k$ is enclosed by the image. Note that, because we are in the weak-lensing regime, this is equivalent to $k\in\interior\source$. Summarizing, at this order of approximation, the convergence is only dictated by the lenses enclosed by the light beam, with
\begin{empheq}[box=\fbox]{equation}\label{eq:convergence}
\kappa = \sum_{k\in\interior\mathcal{S}} \frac{\pi \eps_k^2}{\Omega\e{S}} \ .
\end{empheq}
Equation~\eqref{eq:convergence} agrees with the standard result obtained in the infinitesimal-beam case, where $\kappa$ is related to the projected matter density experienced by the light beam. It was used in \FLUletter to address the so-called Ricci-Weyl dichotomy of gravitational lensing.

Remember that, here, the squared Einstein radii~$\eps_k^2$ are virtually allowed to be negative to account for underdense regions of the Universe, with respect to the background FLRW model. In terms of convergence, it would translate into the possibility of having demagnified lines of sight, where objects appear smaller than they would in the background.

Let us finally comment on the definition of convergence. Here we considered its geometric definition, based on comparing angular sizes, but another standard definition consists in comparing luminous intensities (power per unit area on the detector) as
\begin{equation}
\kappa_I \define \frac{I - I\e{S}}{2 I\e{S}} \ .
\end{equation}
For infinitesimal beams, and if the Universe is perfectly transparent, then $\kappa_I = \kappa$ as a consequence of the distance duality relation, or surface brightness conservation. This property is easily extended to finite beams when the source has a homogeneous surface brightness, but corrections should be expected for realistic inhomogeneous sources~\cite{CFR}.

\subsection{Shear}

Besides convergence, which describes the relative enhancement of their sizes, lensing also affects the shape of images. For infinitesimal sources, these distortions consist in a shear mode---a tiny circular source would appear as a tiny ellipse---encoded in the symmetric traceless part of the amplification matrix~$\mat{\amplification}$. This subsection shows how one can describe and model this effect in the case of extended sources.

\subsubsection{Image quadrupole, complex ellipticity, and shear}

In weak-lensing observations, shear is extracted from the apparent ellipticity of galaxies. A simple estimator for this ellipticity is based on the quadrupole of the image pattern~\cite{1991MNRAS.251..600B}
\begin{equation}\label{eq:quadrupole_def}
\quadrupole_{ab} 
= \frac{\int W[I(\vect{\theta})] \, \theta_a \theta_b \; \dd^2\vect{\theta}}
			{\int W[I(\vect{\theta})] \; \dd^2\vect{\theta}} \ ,
\end{equation}
where $I(\vect{\theta})$ is the image surface brightness in the direction $\vect{\theta}$, and $W$ is a weighting function. In Eq.~\eqref{eq:quadrupole_def}, the coordinate system is chosen to be the $W$-center of the image, defined such that $\int W[I(\vect{\theta})]\vect{\theta} \, \dd^2\vect{\theta} = \vect{0}$. An estimator of the complex ellipticity\footnote{The usual notation for this ellipticity is $\chi$, but we chose to call it $E$ in order to avoid confusion with the comoving radial coordinate.} of the image is then~\cite{1995A&A...294..411S}
\begin{equation}\label{eq:ellipticity_def}
E
\define \frac{2(\quadrupole_{\langle 11\rangle} + \i \quadrupole_{\langle 12\rangle})}
						{\tr \mat{\quadrupole}}
= \frac{\quadrupole_{11}-\quadrupole_{22} + 2\i \quadrupole_{12}}
		{\quadrupole_{11}+\quadrupole_{22}} \ ,
\end{equation}
where angular brackets $\langle ab \rangle$ refer to the traceless part of a matrix. This definition is motivated by the fact that, if the image is an ellipse with semimajor axis~$a$ and semiminor axis~$b$, then $E = (a^2-b^2)(a^2+b^2)^{-1}\ex{2\i\vartheta}$, where $\vartheta$ indicates the direction of the semimajor axis~\cite{2001PhR...340..291B}. Another historically popular estimator also involves the determinant of $\mat{\quadrupole}$~\cite{1997A&A...318..687S}, see Ref.~\cite{2001PhR...340..291B} for a comparison.

In fact, modern techniques for measuring the ellipticity of galaxies, used in the KiloDegree Survey (KiDS)~\cite{Conti:2016gav} and the Dark Energy Survey (DES)~\cite{2016MNRAS.460.2245J} are not based on such estimators, but rather on fitting galaxy models~\cite{2007MNRAS.382..315M} including a disk and a bulge component. This class of methods allows one to better control various sources of noise and systematics. However, since we are here interested in light propagation issues rather than data analysis issues, it seems reasonable to consider Eq.~\eqref{eq:ellipticity_def} a suitable theoretical estimator of image ellipticities.

For an infinitesimal source, the amplification matrix is homogeneous across the image, which implies that the lensing map is effectively linear, $\vect{\beta} = \mat{\amplification}\,\vect{\theta}$. Changing the integration variable~$\vect{\theta}\mapsto\vect{\beta}$ in the definition~\eqref{eq:quadrupole_def} of the image quadrupole, and using surface-brightness conservation, $I(\vect{\theta})=I\e{S}[\vect{\beta}(\vect{\theta})]$, it is straightforward to show that
\begin{equation}\label{eq:quadrupole_transformation}
\vect{\quadrupole}=\transpose{\vect{\amplification}}\vect{\quadrupole}_{\source}\vect{\amplification} \ ,
\end{equation}
where $\vect{\quadrupole}_{\source}$ is the intrinsic quadrupole of the source.
In the weak-lensing regime, it can be shown that $\mat{\amplification}$ is symmetric, as its rotation component turns out to be second order~\cite{Fleury}. Its decomposition into convergence and shear then reads
\begin{equation}
\mat{\amplification}
=
\begin{bmatrix}
1-\kappa-\gamma_1 & \gamma_2 \\
\gamma_2 & 1-\kappa+\gamma_1
\end{bmatrix} \ ,
\end{equation}
so that the transformation law for image ellipticities reads
\begin{align}
E\e{S}
&= E \pac{ 1 + 2\Re\pa{\gamma^* E} } - 2\gamma + \mathcal{O}(\kappa^2, \gamma^2, \kappa\gamma) \\
&= E - 2 \gamma + \ldots
\end{align}
where $E\e{S}$ is the intrinsic ellipticity of the source, and, in the last line, we assumed a quasicircular image ($E \ll 1$). This approximation is clearly not realistic as far as galaxies are concerned, but as we wish to focus on the corrections to shear~$\gamma$, we consider this issue as secondary. We refer the interested reader to \FLUadvanced, where the issue of noncircularity is discussed in detail.

\subsubsection{Extended sources: From the quadrupole to Fourier}

For extended sources, the amplification matrix~$\mat{\amplification}$ is not homogeneous across the image; on the contrary, it can experience significant variations, which prevent one from applying Eq.~\eqref{eq:quadrupole_transformation} for the transformation of the image quadrupole. From now on, we assume that $W(I)$ is a top-hat function with respect to a given brightness threshold~$I\e{c}$, in other words $W=1$ for $I\geq I\e{c}$ (inside the image), and $0$ otherwise. In such conditions the quadrupole reads
\begin{align}
\quadrupole_{ab} 
&= \frac{1}{\Omega} \int_{\interior\image} \theta_a \theta_b \; \dd^2\vect{\theta} \\
&= \frac{1}{4\Omega} \int_{\interior\image} \pd{(\theta_a \theta_b \theta_c)}{\theta_c} \; \dd^2\vect{\theta} 
	\label{eq:rewriting_quadrupole_1}\\
&= \frac{1}{4\Omega} \int_{\image} \theta_a \theta_b \, \det(\vect{\theta},\dd\vect{\theta})
	\label{eq:rewriting_quadrupole_2}\\
&= \frac{1}{4\Omega} \int_0^{2\pi} \theta^4 e_a e_b \; \dd\psi
\end{align}
where from Eq.~\eqref{eq:rewriting_quadrupole_1} to Eq.~\eqref{eq:rewriting_quadrupole_2} we used Stokes' theorem, and in the last line we introduced polar coordinates, $\vect{\theta} = \theta \vect{e}$, with $\vect{e}=(\cos\psi, \sin\psi)$. The trace of $\vect{\quadrupole}$ and its traceless component are then easily found to read
\begin{align}
\tr\mat{\quadrupole}
&= \frac{1}{4\Omega} \int_{0}^{2\pi} \theta^4 \; \dd\psi \ , \\
2(\quadrupole_{\langle 11\rangle}+\i \quadrupole_{\langle 12\rangle})
&= \frac{1}{4\Omega} \int_{0}^{2\pi} \theta^4 \ex{2\i \psi} \; \dd\psi \ ,
\end{align}
which together imply
\begin{equation}\label{eq:ellipticity_Fourier}
E = \frac{\int_0^{2\pi} \theta^4 \ex{2\i\psi} \; \dd\psi}{\int_0^{2\pi} \theta^4 \; \dd\psi} \ .
\end{equation}
Ellipticity then appears as a ratio of Fourier modes of the periodic function~$\psi\mapsto\theta^4(\psi)$, which describes the contour~$\image$ of the image.

\subsubsection{Exterior and interior shear}

The last step in the calculation of $E$ consists in reintroducing the complex notation for sources and images. If $\cplx{\theta}=\theta\ex{\i\psi}$ is the complex position of an image point, then $\cplx{\beta} = \beta\ex{\i\ph}$ is the associated source point. Note that the angles~$\psi$ and $\ph$ are generally different, because the difference
\begin{align}
\delta\cplx{\theta} &\define \cplx{\theta} - \cplx{\beta} \\
&= \sum_k \frac{\eps_k^2}{\cplx{\theta}^* - \cplx{\lens}^*_k} \\
&= \sum_k \frac{\eps_k^2}{\cplx{\beta}^* - \cplx{\lens}^*_k} + \mathcal{O}(\eps^4)
	\label{eq:delta_theta}
\end{align}
is not necessarily aligned with $\cplx{\beta}$.

At lowest order in $\delta\cplx{\theta}\sim \eps^2$,
\begin{equation}
\theta^4
= |\cplx{\theta}|^4 = \beta^4 + 2\beta^3 \pa{ \ex{-\i\ph} \delta\cplx{\theta} +  \ex{\i\ph}\delta\cplx{\theta}^*}
 + \mathcal{O}(\eps^4) \ ,
\end{equation}
and if we assume, for simplicity, that the source is circular ($\beta=\cst$), then the complex ellipticity simply reads
\begin{equation}\label{eq:ellipticity_delta_theta}
E = \frac{1}{\pi\beta} \pac{ \int_0^{2\pi} \ex{\i\ph} \delta\cplx{\theta} \; \dd\ph
										+ \pa{\int_0^{2\pi} \ex{-3\i\ph} \delta\cplx{\theta} \; \dd\ph}^* } + \mathcal{O}(\eps^4) \ .
\end{equation}
Note that, in Eq.~\eqref{eq:ellipticity_delta_theta}, we replaced the integration over the angular position of the image~$\psi$ by an integration over the angular position of the source~$\beta$. This is justified by the fact that, for a given source-image couple, $\psi-\ph = \Im(\cplx{\beta}^{-1}\delta\cplx{\theta}) = \mathcal{O}(\eps^2)$, and hence the difference between $\dd\psi$ and $\dd\ph$ would yield a term $\mathcal{O}(\eps^4)$ in the integrals of $\delta\cplx{\theta}$. Finally, the above is easily generalized to the case of a quasicircular source, writing $\cplx{\theta} = \beta\ex{\i\ph} + \delta\cplx{\beta} + \delta\cplx{\theta}$. This just adds an intrinsic ellipticity term~$E\e{S}$ to Eq.~\eqref{eq:ellipticity_delta_theta}. Identifying with the infinitesimal-source case, we can thus consider that Eq.~\eqref{eq:ellipticity_delta_theta} is, in fact, the expression of $2\gamma$. Things are more involved when the source cannot be considered quasicircular, in particular the shear becomes entangled with the intrinsic ellipticity of the source; see \FLUadvanced for a detailed discussion.

We now proceed with calculating the integrals of Eq.~\eqref{eq:ellipticity_delta_theta}. Again, these are elegantly dealt with using the residue theorem. First define the Fourier modes of $\delta\cplx{\theta}(\ph)$,
\begin{align}
\delta\cplx{\theta}_n &\define \frac{1}{2\pi}\int_0^{2\pi} \ex{-\i(n+1)\ph} \delta\cplx{\theta}(\ph) \; \dd\ph \ , \\
\delta\cplx{\theta}(\ph) &= \sum_{n\in\mathbb{Z}} \delta\cplx{\theta}_n \, \ex{\i(n+1)\ph}, 
\end{align}
such that
\begin{equation}
\gamma = \frac{1}{\beta} \pa{\delta\cplx{\theta}_{-2}+\delta\cplx{\theta}_2^*} \ ,
\end{equation}
then introduce the complex lens equation~\eqref{eq:delta_theta} at lowest order in $\eps^2$, change variable from $\ph$ to $-\ph$, and transform the angular integral into a complex integral with $\cplx{\beta} =\beta\ex{\i\ph}$,
\begin{align}
\delta\cplx{\theta}_n
&= \sum_k \frac{\eps_k^2}{2\pi} 
		\int_0^{2\pi} \frac{\ex{-\i(n+1)\ph}}{\cplx{\beta}^*-\cplx{\lambda}_k^*} \; \dd\ph \\
&= \sum_k \frac{\eps_k^2}{2\pi}
		\int_0^{2\pi} \frac{\ex{\i(n+1)\ph}}{\cplx{\beta}-\cplx{\lambda}_k^*} \; \dd\ph \\
&= \sum_k \frac{\eps_k^2}{2\i\pi\beta^{n+1}}
		\ointctrclockwise_{\source} \frac{\cplx{\beta}^n \, \dd \cplx{\beta}}{\cplx{\beta} - \cplx{\lambda}_k^*} \ .
		\label{eq:complex_integrals_shear}
\end{align}
If $n\geq 0$, the integrand of Eq.~\eqref{eq:complex_integrals_shear} only has a pole at $\cplx{\beta}=\cplx{\lambda}_k^*$. The associated residue, $(\cplx{\lambda}_k^*)^n$, only counts if $\cplx{\lambda}_k \in \interior\source$, and hence
\begin{equation}
\forall n\geq 0 \qquad
\delta\cplx{\theta}_{n} = \sum_{k\in\interior\source}
														\frac{\eps_k^2}{\beta} \pa{\frac{\cplx{\lambda}_k^*}{\beta}}^n \ .
\end{equation}
If $n<0$, the integrand of Eq.~\eqref{eq:complex_integrals_shear} has an additional pole at $\cplx{\beta}=0$, and the associated residue turns out to be exactly opposite to the residue at $\cplx{\beta} = \cplx{\lambda}_k^*$. These residues thus compensate if $\cplx{\lambda}_k\in\interior\source$, and nonzero terms now come from $\cplx{\lambda}_{k}\in\exterior\source$, where $\exterior\source$ denotes the set of the complex plane consisting of points outside the contour $\source$,
\begin{equation}
\forall n<0 \qquad
\delta\cplx{\theta}_{n} = -\sum_{k\in\exterior\source}
														\frac{\eps_k^2}{\beta} \pa{\frac{\cplx{\lambda}_k^*}{\beta}}^n \ .
\end{equation}

\begin{figure}[h!]
\centering
\includegraphics[width=\columnwidth]{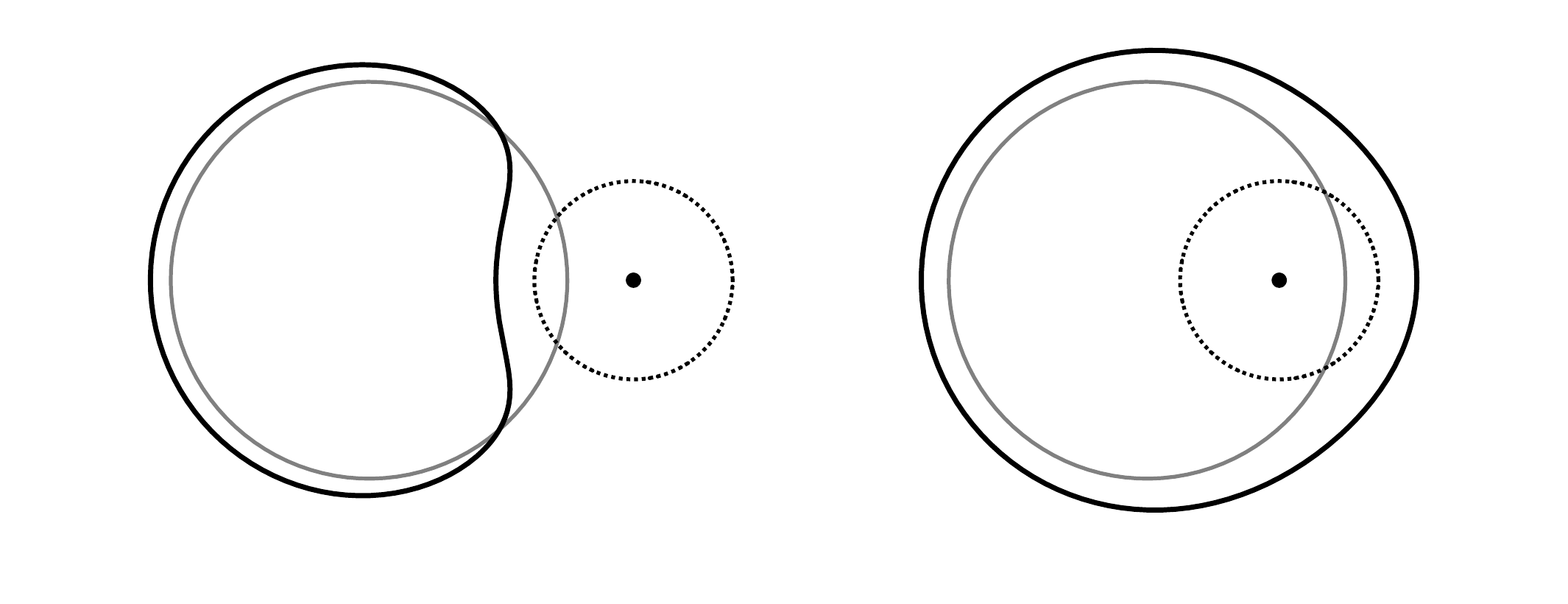}
\caption{Comparison of the distortion effects of an exterior lens (left) and an interior lens (right), represented by black disks. Grey lines indicate the contour~$\source$ of the circular source, with radius~$\beta$; thick solid lines are the contour~$\image$ of their images; and dotted lines represent the Einstein radius of the lens, chosen as $\eps=\beta/2$ here. To make the effect more visible, we went far beyond the weak-lensing approximation, with $|\vect{\beta} - \vect{\lambda}|\e{min}=2\eps/3 \not\gg \eps$.}
\label{fig:inside_out}
\end{figure}

Applying these results to $n=\pm 2$, we conclude that the shear of finite sources picks up two different contributions:
\begin{empheq}[box=\fbox]{equation}\label{eq:shear_ext_int}
\gamma =
\underbrace{ - \sum_{k\in\exterior\source} \pa{\frac{\eps_k}{\cplx{\lens}_k^*}}^2 }_{\gamma\e{ext}}
+ \underbrace{ \sum_{k\in\interior\source} \pa{\frac{\pi\eps_k \cplx{\lens}_k}{\Omega\e{S}}}^2 }_{\gamma\e{int}} \ .
\end{empheq}
The first term, $\gamma\e{ext}$, is due to the lenses located outside the beam; the impact of a given lens in this term decreases proportionally to the inverse square of its distance to the center of the source, which is the standard behavior of the weak-lensing shear. The second term, $\gamma\e{int}$, is a new contribution due to the lenses enclosed by the beam. By definition, such a contribution cannot be accounted for by a model where light beams and sources are infinitesimal. Figure~\ref{fig:inside_out} illustrates the distortions of a circular source due to an interior or an exterior lens. Note that shear occurs in orthogonal directions depending on whether the lens is inside or outside~$\source$. This corresponds to the opposite signs of $\gamma\e{int}$ and $\gamma\e{ext}$ in Eq.~\eqref{eq:shear_ext_int}.

One could be worried about the fact that~$\gamma\e{int}$ seems to diverge like $\beta^{-4}$ as $\beta\rightarrow 0$. Fortunately this divergence is only apparent. A closer examination of Eq.~\eqref{eq:shear_ext_int} already shows that, due to the presence of $\cplx{\lambda}_k<\beta$ in the numerator of $\gamma\e{int}$, this term would at most diverge like $\beta^{-2}$ rather than $\beta^{-4}$. Furthermore, since only interior lenses contribute to $\gamma\e{int}$, the probability that a random lens indeed contributes to this sum is expected to go like $\Omega\e{S}=\pi \beta^2$, which finally compensates the divergence.

\subsection{Violation of the Kaiser-Squires relation?}

In \FLUletter, where the above results were first exposed, we compared the statistical variance of convergence and shear, in a Universe randomly filled with point lenses, with no correlations between their positions (Poisson distribution). We found that, in that case, the contribution of interior lenses to $\ev[2]{|\gamma|^2}$ was statistically significant---namely, equal to a third of the standard exterior contribution. As a consequence,
\begin{equation}\label{eq:violation_KS}
\ev[2]{|\gamma|^2} = \frac{4}{3} \, \ev[2]{\kappa^2} \ ,
\end{equation}
which is in violent contradiction with the standard results for infinitesimal beams
\begin{equation}
\ev[2]{|\gamma|^2} = \ev[2]{\kappa^2} \ ,
\end{equation}
following from the Kaiser-Squires equality between convergence and shear power spectra~\cite{1993ApJ...404..441K}. Although the model matter distribution (randomly arranged point masses) yielding Eq.~\eqref{eq:violation_KS} is unrealistic, this result suggests that there may be systematic biases in cosmic shear measurements, which must be investigated, given the growing importance of weak gravitational lensing in current and future cosmological surveys.

The remainder of this article is dedicated to this task. We will show that, as already suspected in \FLUletter, Eq.~\eqref{eq:violation_KS} highly overestimates the amplitude of finite-beam corrections for realistic cosmic shear measurements. We will also explain the reasons of this overestimation.

\section{Cosmic weak lensing}
\label{sec:cosmic_weak_lensing}

\subsection{From discrete to continuous}
\label{sec:discrete_to_continuous}

In cosmology, matter is modeled as a (set of) fluid(s), or at least a continuous medium, described by its density field~$\rho$, rather than a set of particles with their individual mass and position. In order to apply the results of Sec.~\ref{sec:finite-beam_formalism} to cosmic weak lensing, the first step thus consists in reformulating it in terms of such a continuous medium. Fortunately, such a step turns out to be quite straightforward, because both expressions of convergence~$\kappa$~\eqref{eq:convergence} and shear~$\gamma$~\eqref{eq:shear_ext_int} appear as sums of terms~$\propto \eps_k^2 \propto m_k$. Going from a discrete model of matter distribution to a continuous model will thus consist in turning sums into integrals as
\begin{equation}
\sum_k m_k \, (\ldots) \rightarrow \int \dd^3 m \, (\ldots) = \int \delta\rho \, \dd^3 V \, (\ldots) \ ,
\end{equation}
where $\delta\rho \define \rho - \bar{\rho}$ is the density relative to the FLRW background. This correspondence between discrete and continuous descriptions of matter, involving $\delta\rho$ instead of $\rho$, is the reason why we allowed the masses~$m_k$ to be negative in Sec.~\ref{sec:finite-beam_formalism}.

Substituting the expression~\eqref{eq:Einstein_radius} of $\eps_k$, we find that convergence, for example, reads
\begin{equation}\label{eq:convergence_continuous}
\kappa = \frac{4\pi G}{\Omega\e{S}} \int_{\beam} \delta\rho \, \dd^3 V \; 
				(1+z) \frac{f_K(\chi\e{S}-\chi)}{f_K(\chi\e{S})f_K(\chi)} \ ,
\end{equation}
where integration is performed over the region of space covered by the (background) light beam~$\beam$. If we introduce the density contrast~$\delta \define \delta\rho/\bar{\rho}$, and assume that $\bar{\rho}$ is dominated by nonrelativistic matter (apart from dark energy), we obtain
\begin{equation}
\delta\rho \, \dd^3 V = \bar{\rho}  \delta\, \dd^3 V = \bar{\rho}_0 \delta \, \dd^3 V_0
\end{equation}
where a subscript~$0$ indicates the value of a background quantity today. The volume element~$\dd^3 V_0$, in particular, can be expressed in terms of the spatial coordinates as
\begin{equation}
\dd^3 V_0 \approx \dd\chi \, f_K^2(\chi) \dd\lens \, \lens \dd\phi
\end{equation}
in the flat-sky approximation, and where we assumed~$a_0=1$. We used the notation $\vect{\lens} = \lens(\cos\phi, \sin\phi)$ in order to keep track of the meaning of this variable, which appears in the argument of $\delta\rho$, and hence locates the position of the lenses. If $\vect{\alpha}$ is the center of the source, we finally get
\begin{multline}\label{eq:convergence_fixed_z}
\kappa(\chi\e{S}, \vect{\alpha})
= 4\pi G\bar{\rho}_0 \int_0^{\chi\e{S}} \dd\chi \; (1+z) \, \frac{f_K(\chi\e{S}-\chi)f_K(\chi)}{f_K(\chi\e{S})} \\
																				 \times \bar{\delta}_\beam(\eta_0-\chi, \chi, \vect{\alpha}) \ ,
\end{multline}
where $\bar{\delta}_\beam$ represents the density contrast averaged over the beam's cross section,
\begin{equation}
\bar{\delta}_\beam(\eta,\chi,\vect{\alpha})
\define \frac{1}{\Omega\e{S}} \int_{\interior\source} \dd^2\vect{\lambda} \; \delta(\eta, \chi, \vect{\alpha} + \vect{\lens}) \ ,
\end{equation}
as expected from the property that light beams are smoothing out the matter distribution that they encounter~\cite{Fleury:2017owg}. From now on, we restrict our analysis to circular sources with unlensed radius~$\beta$, so that
\begin{equation}
\bar{\delta}_\beam(\eta,\chi,\vect{\alpha})
\define \frac{1}{\pi \beta^2}
	\int_0^{\beta} \lens\,\dd\lens \int_0^{2\pi} \dd\phi \; \delta(\eta, \chi, \vect{\alpha} + \vect{\lens}) \ .
\end{equation}

\subsection{Effective convergence}

For surveys measuring weak-lensing signals by stacking many sources in various redshift bins, it is customary to introduce the notions of effective convergence and effective shear, by averaging those quantities over the set of sources. Here, since convergence depends not only on the distance~$\chi_*$ to the source, but also on the angular radius~$\beta$ of this source, averaging has to be taken relatively to both parameters. Let us denote $p(\beta, \chi_*)$ the associated joint probability density function (PDF), then the effective convergence is defined as
\begin{equation}\label{eq:effective_convergence_def}
\kappa\e{eff}(\vect{\alpha})
\define \int_0^{\chi\e{H}} \dd\chi_* \, \dd\beta \; p(\beta, \chi_*) \, \kappa(\chi_*, \vect{\alpha}) \ ,
\end{equation}
where $\chi\e{H}$ is the comoving radius of the particle horizon.

A reasonable simplifying assumption consists in considering that the intrinsic physical radius~$r$ of a source is independent of its distance from the observer. If a source at $\chi_*$ is comoving with the cosmological background, then $r=f_K(\chi_*)\beta/(1+z_*)$, and hence
\begin{align}
p(\beta, \chi_*)
&=  p_\beta(\beta|\chi_*) p_\chi(\chi_*)\\
&= \frac{f_K(\chi_*)}{1+z_*} \, p_r\pac{ \frac{f_K(\chi_*)\beta}{1+z_*}} p_\chi(\chi_*) \ ,
\label{eq:joint_PDF}
\end{align}
where $p_r$ is the PDF of the intrinsic radius (size) of the sources.

Inserting the expression~\eqref{eq:convergence_fixed_z} of $\kappa$ into Eq.~\eqref{eq:effective_convergence_def}, and inverting integration order as $\int_0^{\chi\e{H}}\dd\chi_*\int_0^{\chi_*} \dd\chi = \int_0^{\chi\e{H}}\dd\chi\int_\chi^{\chi\e{H}}\dd\chi_*$, we can put the effective convergence under a more familiar form,
\begin{multline}\label{eq:effective_convergence_result}
\kappa\e{eff}(\vect{\alpha})
= 4\pi G \bar{\rho}_0 \int_0^\infty \dd\beta \int_0^{\chi\e{H}} \dd\chi \, (1+z) f_K(\chi) \\
	\times q(\beta, \chi) \, \bar{\delta}_\beam(\eta_0-\chi, \chi, \vect{\alpha}) \ ,
\end{multline}
with the weighting function
\begin{equation}
q(\beta, \chi)
\define \int_\chi^{\chi\e{H}}
			\dd\chi_* \;  p(\beta, \chi_*) \, \frac{f_K(\chi_*-\chi)}{f_K(\chi_*)} \ .
\end{equation}
Note that for infinitesimal sources, $p_r(r) = \delta\e{D}(r)$, one recovers the standard result.

\subsection{Effective shear}

Let us proceed with the same kind of calculations for shear. We now have two terms, respectively associated with the contributions of interior and exterior lenses. With a continuous description of matter, we thus have
\begin{align}
\gamma\e{ext} 
&=
-4G \int_{\mathbb{R}^3\setminus\beam} \frac{\delta\rho \, \dd^3 V}{(\cplx{\lens}^*)^2} \; 
\frac{(1+z)f_K(\chi\e{S}-\chi)}{f_K(\chi\e{S})f_K(\chi)} \ , \\
\gamma\e{int} 
&=
4G \int_{\beam} \frac{\pi^2\cplx{\lambda}^2\delta\rho \, \dd^3 V}{\Omega\e{S}^2} \; 
\frac{(1+z)f_K(\chi\e{S}-\chi)}{f_K(\chi\e{S})f_K(\chi)}  \ ,
\end{align}
quite similarly to Eq.~\eqref{eq:convergence_continuous}. Then, following the same lines as in the case of convergence, we find that the effective shear takes the same form as Eq.~\eqref{eq:effective_convergence_result}
\begin{multline}
\gamma\e{eff}(\vect{\alpha})
=
4\pi G \bar{\rho}_0 \int_0^\infty \dd\beta \int_0^{\chi\e{H}} \dd\chi \, (1+z) f_K(\chi) \\
	\times q(\beta, \chi) \, (\Gamma_\beam*\delta)(\eta_0-\chi, \chi, \vect{\alpha}) \ ,
\end{multline}
except that the beam-averaged density contrast is replaced by a convolution product
\begin{equation}
(\Gamma_\beam*\delta)(\eta, \chi, \vect{\alpha})
\define
\int_{\mathbb{R}^2} \frac{\dd^2\vect{\lambda}}{\Omega\e{S}} \; 
											\Gamma_\beam(\vect{\lens}) \, \delta(\eta_0-\chi, \chi, \vect{\alpha}+\vect{\lens})
\end{equation}
between $\delta$ and the shear kernel $\Gamma_\beam = \Gamma_\beam\h{ext} + \Gamma_\beam\h{int}$, with
\begin{align}
\Gamma_\beam\h{ext}(\vect{\lambda})
&\define -\Theta(\lambda-\beta) \pa{ \frac{\beta}{\lens}}^2 \ex{2\i\phi} \ , \\
\Gamma_\beam\h{int}(\vect{\lambda})
&\define \Theta(\beta-\lambda) \pa{ \frac{\lens}{\beta}}^2 \ex{2\i\phi} \ ,
\end{align}
and where $\Theta$ denotes the Heaviside function. Recall that we are dealing with circular sources, and that $\vect{\lens}$ span the position of lenses, therefore $\Theta(\lambda-\beta)$ selects exterior lenses, while $\Theta(\beta-\lens)$ selects interior lenses.

\section{Two-point correlations}
\label{sec:two-point_correlations}

In the previous section, we derived the expression of the effective convergence and shear for extended sources, as functions of the density contrast field. Just as the standard infinitesimal-beam case, we can now deduce the corresponding angular correlation functions and power spectra.

\subsection{Convergence}
\label{sec:convergence}

The correlation function of the convergence~$\xi_\kappa$ and its angular power spectrum~$P_\kappa$ are defined and related as
\begin{align}
\ev{ \kappa\e{eff}(\vect{\alpha}_1) \kappa\e{eff}(\vect{\alpha}_2) }
&= \xi_\kappa(|\vect{\alpha}_1-\vect{\alpha}_2|) \\
&= \int_{\mathbb{R}^2} \frac{\dd^2\vect{\ell}}{(2\pi)^2} \,
		\ex{\i\vect{\ell}\cdot(\vect{\alpha}_1-\vect{\alpha}_2)} \, P_\kappa(\ell)\label{eq:identify_convergence_spectrum}\\
&= \int_0^\infty \frac{\dd\ell}{2\pi} \, \ell J_0(\ell|\vect{\alpha}_1-\vect{\alpha}_2|) P_\kappa(\ell),
\end{align}
where angular brackets~$\ev{\ldots}$ denote ensemble average, and $J_n$ is the $n$th-order Bessel function. Here we adopted the flat-sky notation~$P_\kappa(\ell)$ for the angular power spectrum; for small scales ($\ell\gg 1$), which is the regime that we are interested in, it corresponds to the multipole~$C_\ell^\kappa$.

Following the standard procedure, we insert the expression~\eqref{eq:effective_convergence_result} of $\kappa\e{eff}$ into the definition of $\xi_\kappa$, and make the expectation value enter the integral; however, contrary to the infinitesimal-beam case, the resulting integrand is not $\propto \ev{\delta \delta}$, but $\propto\ev[2]{\bar{\delta}_{\beam_1}\bar{\delta}_{\beam_2}}$, where $\beam_1$ and $\beam_2$ are the two beams which are correlated. Assuming that the beam sizes are independent of the distribution of matter they encounter, and following on the latter, we thus have to compute
\begin{multline}
\ev{\bar{\delta}_{\beam_1}\bar{\delta}_{\beam_2}}
= \int_{\beam_1} \frac{\dd^2\vect{\lens}_1}{\pi\beta_1^2} \int_{\beam_2} \frac{\dd^2\vect{\lens}_2}{\pi\beta_2^2} \;
		\Big\langle\delta(\eta_0-\chi_1,\chi_1,\vect{\alpha}_1+\vect{\lambda}_1) \\
		\times\delta(\eta_0-\chi_2,\chi_2,\vect{\alpha}_2+\vect{\lambda}_2) \Big\rangle \ .
\end{multline}
This is nothing but a filtered version of the matter correlation function, from which scales smaller than $\beam_1, \beam_2$ are removed. More quantitatively, inserting the Fourier transform of the density contrast and the associated power spectrum, with the convention
\begin{equation}
\ev{\delta(\eta_1, \vect{k}_1) \delta(\eta_2, \vect{k}_2)}
= (2\pi)^3 \delta\e{D}(\vect{k}_1+\vect{k}_2) \, P_\delta(\eta_1, \eta_2, k_1) \ ,
\end{equation}
and using Limber's approximation~\cite{Kaiser:1991qi}, we find
\begin{multline}
\ev{\bar{\delta}_{\beam_1}\bar{\delta}_{\beam_2}}
\approx
\frac{\delta\e{D}(\chi_1-\chi_2)}{f_K(\chi_1)} 
 \int_{\beam_1} \frac{\dd^2\vect{\lens}_1}{\pi\beta_1^2} \int_{\beam_2} \frac{\dd^2\vect{\lens}_2}{\pi\beta_2^2}
 \int_{\mathbb{R}^2} \frac{\dd^2\vect{\ell}}{(2\pi)^2} \\
 \times \ex{\i\vect{\ell}\cdot(\vect{\alpha}_1-\vect{\alpha}_2)} \ex{\i\vect{\ell}\cdot(\vect{\lens}_1-\vect{\lens}_2)}
 P_\delta\pac{\eta_0-\chi_1, \frac{\ell}{f_K(\chi_1)}} .
\end{multline}
Integration over $\vect{\lambda}_1$, $\vect{\lambda}_2$ yields two Bessel functions~$J_1(\ell\beta_1)$, $J_1(\ell\beta_2)$, and by comparing with Eq.~\eqref{eq:identify_convergence_spectrum}, we finally read

\vspace*{1mm}

\noindent\fbox{
\begin{minipage}{0.95\columnwidth}
\vspace*{-0.3cm}
\begin{multline}\label{eq:convergence_power_spectrum}
P_\kappa(\ell) =
\pa{\frac{3}{2} H_0^2\Omega\e{m}}^2 \int_0^{\chi\e{H}} \dd\chi \; (1+z)^2 \, \bar{q}_\kappa^2(\ell, \chi) \\
																			\times P_\delta\pac{\eta_0-\chi, \frac{\ell}{f_K(\chi)} }
\end{multline}
with the modified lensing weight
\begin{equation}\label{eq:kernel_convergence}
\bar{q}_\kappa
\define \int_0^\infty \dd\beta \; \frac{2 J_1(\ell\beta)}{\ell\beta}
				\int_\chi^{\chi\e{H}} \dd\chi_* \; p(\beta, \chi_*) \, \frac{f_K(\chi_*-\chi)}{f_K(\chi_*)} \ .
\end{equation}
\end{minipage}
}

\vspace*{1mm}

In Eq.~\eqref{eq:convergence_power_spectrum}, the extension of sources is fully encoded in the Bessel term $2J_1(\ell\beta)/(\ell\beta)$ of the lensing weighting function~$\bar{q}_\kappa$. The infinitesimal-beam result is naturally recovered for $p(\beta,\chi)\propto\delta\e{D}(\beta)$, because $\lim{x}{0} 2J_1(x)/x = 1$, so that
\begin{equation}
\bar{q}_\kappa \rightarrow 
\int_\chi^{\chi\e{H}} \dd\chi_* \; p_\chi(\chi_*) \, \frac{f_K(\chi_*-\chi)}{f_K(\chi_*)} \ .
\end{equation}

It is also instructive to consider the special case where all sources are located at the same comoving distance~$\chi$, and have the same intrinsic size~$r$---in that case, they also have the same unlensed angular radius~$\beta=(1+z) r/f_K(\chi)$, and the result is
\begin{equation}\label{eq:damping_convergence_simple}
P_\kappa(\ell) = P_\kappa^{0}(\ell) \times \pac{\frac{2 J_1(\ell\beta)}{\ell\beta} }^2,
\end{equation}
where $P_\kappa^{0}(\ell)$ denotes the infinitesimal-beam convergence power spectrum. The damping factor is depicted in Fig.~\ref{fig:damping_power_spectrum}.

\subsection{Shear}
\label{sec:shear}

The shear being a complex quantity, there is more freedom in the definition of its correlations. Two real correlation functions are usually defined, namely
\begin{align}
\xi_+(|\vect{\alpha}_1-\vect{\alpha}_2|)
&\define \ev{ \gamma\e{eff}(\vect{\alpha}_1) \gamma\e{eff}^*(\vect{\alpha}_2) } \ ,\\
\xi_-(|\vect{\alpha}_1-\vect{\alpha}_2|)
&\define \ev{ \gamma\e{eff}(\vect{\alpha}_1) \gamma\e{eff}(\vect{\alpha}_2) }
				\ex{-4\i \phi_{\vect{\alpha}_1-\vect{\alpha}_2}} \ ,
\label{eq:xi_minus_def}
\end{align}
where $\phi_{\vect{\alpha}_1-\vect{\alpha}_2}$ is the polar angle of $\vect{\alpha}_1-\vect{\alpha}_2$. The shear power spectrum is defined as the Fourier transform of $\xi_+$, and it is related to $\xi_-$ as
\begin{align}
\xi_+(\alpha) &= \int_0^\infty \frac{\dd\ell}{2\pi} \, \ell J_0(\ell\alpha) P_\gamma(\ell) \ , \\
\xi_-(\alpha) &= \int_0^\infty \frac{\dd\ell}{2\pi} \, \ell J_4(\ell\alpha) P_\gamma(\ell) \ .
\label{eq:xi_minus_expr}
\end{align}

The calculation of $P_\gamma$ follows the same lines as that of $P_\kappa$, except that~$\bar{\delta}_\beam$ must be replaced by the convolution product $\Gamma_\beam * \delta$. This also generates Bessel terms, and the final result reads

\vspace{0.2cm}

\noindent\fbox{
\begin{minipage}{0.95\columnwidth}
\vspace*{-0.4cm}
\begin{multline}\label{eq:shear_power_spectrum}
P_\gamma(\ell) =
\pa{\frac{3}{2} H_0^2 \Omega\e{m}}^2 \int_0^{\chi\e{H}} \dd\chi \; (1+z)^2 \, \bar{q}_\gamma^2(\ell, \chi) \\
																			\times P_\delta\pac{\eta_0-\chi, \frac{\ell}{f_K(\chi)} }
\end{multline}
with
\begin{equation}\label{eq:kernel_shear}
\bar{q}_\gamma
\define \int_0^\infty \dd\beta \; \frac{4 J_2'(\ell\beta)}{\ell\beta} 
				\int_\chi^{\chi\e{H}} \dd\chi_* \; 
				p(\beta, \chi_*) \, \frac{f_K(\chi_*-\chi)}{f_K(\chi_*)} \ ,
\end{equation}
\end{minipage}
}

\vspace{0.2cm}

\noindent where we used $J_1(x)-J_3(x)=2J_2'(x)$.

\begin{figure}[h!]
\centering
\includegraphics[width=\columnwidth]{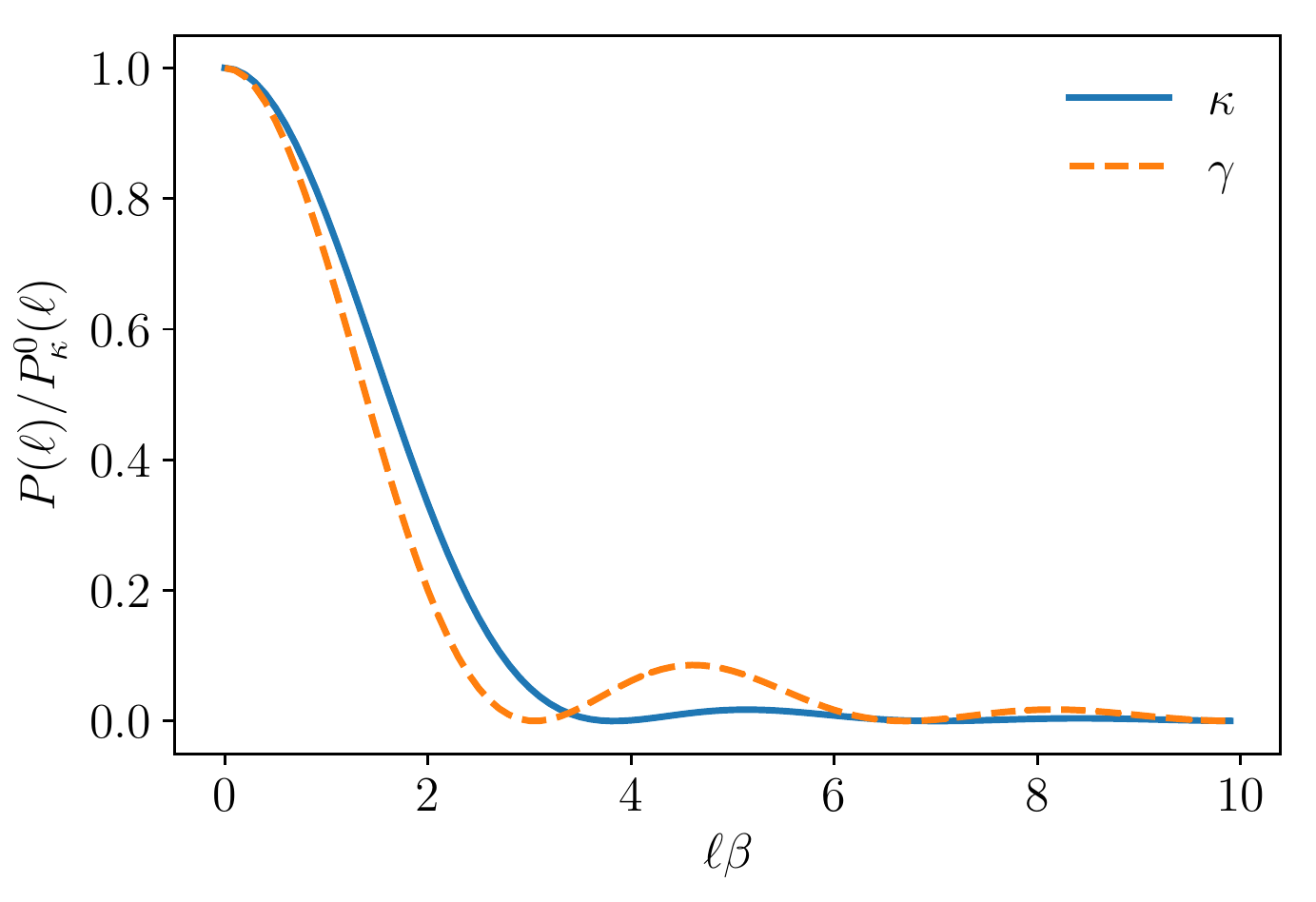}
\caption{Finite-beam damping of the convergence and shear power spectra, $P_\kappa(\ell), P_\gamma(\ell)$ with respect to their infinitesimal beam counterpart~$P^0_\kappa(\ell)=P^0_\gamma(\ell)$, if the sources are all located at the same redshift, and with the same intrinsic size; $\beta$ is their unlensed angular radius.}
\label{fig:damping_power_spectrum}
\end{figure}

Just like for convergence, it is interesting to consider the special case of sources all located at the same redshift, and with the same intrinsic size. In this case, the finite-beam power spectrum is simply damped with respect to the standard case, $P_\gamma^0(\ell) = P_\kappa^0(\ell)$, as
\begin{equation}\label{eq:damping_shear_simple}
P_\gamma(\ell) = P_\kappa^0(\ell) \times \pac{ \frac{4 J_2'(\ell\beta)}{\ell\beta} }^2 ,
\end{equation}
which is depicted in Fig.~\ref{fig:damping_power_spectrum}.

A number of qualitative comments can already be made from these results. On the one hand, it shows that, for extended sources, there is indeed a violation of the Kaiser-Squires equality between the convergence and shear power spectra. As illustrated in Fig.~\ref{fig:damping_power_spectrum}, these effects start to play an important role when $\ell\beta\sim 1$. The typical angular size of a galaxy at $z\approx 0.5$ is $\beta\sim 1" \sim 5\times 10^{-6}\U{rad}$, which corresponds to $\ell \sim 2\times 10^{5}$. This is far beyond the regime probed by current and future lensing surveys. For instance, the multipoles analyzed in KiDS range from $\ell = 76$ to $\ell = 1310$~\cite{2017MNRAS.471.4412K}. In that regime, the Bessel damping terms can be expanded, so that
\begin{equation}
\frac{P_\kappa^0(\ell) - P_\gamma(\ell)}{P_\kappa^0(\ell)} \approx \frac{(\ell\beta)^2}{3} \sim 10^{-5} \ ,
\end{equation}
which is a negligible systematic effect. In the same regime,
\begin{equation}
\frac{P_\kappa^0(\ell) - P_\kappa(\ell)}{P_\kappa^0(\ell)} \approx \frac{(\ell\beta)^2}{4} \ ,
\end{equation}
so that $\gamma$ is more damped than $\kappa$ for $\ell\beta\ll 1$ as we could already see in Fig.~\ref{fig:damping_power_spectrum}. A possible physical interpretation of this property is the presence of anticorrelation due to the situations where a lens is located inside one beam, but outside the other, producing shear effects of opposite signs. Note however that the difference between $P_\kappa$ and $P_\gamma$ flips for $\ell\beta>\pi$. This will turn out to be important to understand the origin of the 4/3 factor between $\ev[1]{\kappa^2}$ and $\ev[1]{|\gamma|^2}$ found in \FLUletter.

\subsection{Special case: Poisson distribution}
\label{sec:4/3}

In \FLUletter, we considered a static universe randomly filled with point lenses, with no correlation between their positions (Poisson distribution). In this case, the correlation function of the density contrast is a Dirac delta, and the associated power spectrum is a constant, $P_\delta(\eta, k)=\cst$. For identical sources that are all located at the same redshift, we thus find, in this special case,
\begin{align}
\xi_\kappa(\alpha)
&= P_\delta \int_0^\infty \frac{\dd\ell}{2\pi} \, \ell J_0(\ell\alpha) \pac{\frac{2 J_1(\ell\beta)}{\ell\beta}}^2 , \\
\xi_+(\alpha)
&= P_\delta \int_0^\infty \frac{\dd\ell}{2\pi} \, \ell J_0(\ell\alpha) \pac{\frac{4 J_2'(\ell\beta)}{\ell\beta}}^2 .
\end{align}

The variance of convergence and shear is obtained for $\alpha=0$, which implies $J_0(\ell\alpha)=1$, so that
\begin{align}\label{eq:variance_ratio}
\frac{\ev[2]{|\gamma|^2}}{\ev{\kappa^2}}
= \frac{\int_0^\infty \dd x \, [2 J_2'(x)]^2/x}{\int_0^\infty \dd x \, [J_1(x)]^2/x} = \frac{4}{3} \ ,
\end{align}
regardless of the apparent size $\beta$ of the sources. The dependence in $\beta$ is removed by the scale invariance of the Poisson distribution. This explains the result of \FLUletter, which could have seemed to be in contradiction with the present article.

In our Universe, the matter power spectrum is not scale independent. In particular, it decreases on small scales because the shorter a mode, the more time it has been sub-Hubble during the radiation era, and therefore the more damped by acoustic effects. As a result, the integrals of Eq.~\eqref{eq:variance_ratio} should be equipped with a kernel that decays as an inverse power law for $x=\ell\beta>1$, where the difference between $J_2'$ and $J_1$ is most important---see Fig.~\ref{fig:damping_power_spectrum}. Therefore, only little power is expected to be affected by finite-beam effects in the real Universe. As will be confirmed more quantitatively in the next subsection, the simplifying assumption of Poisson-distributed matter in \FLUletter led to significant overestimation of those effects.

\subsection{Quantitative finite-beam corrections}
\label{sec:quantitative}

\begin{figure}[h!]
\centering
\includegraphics[width=\columnwidth]{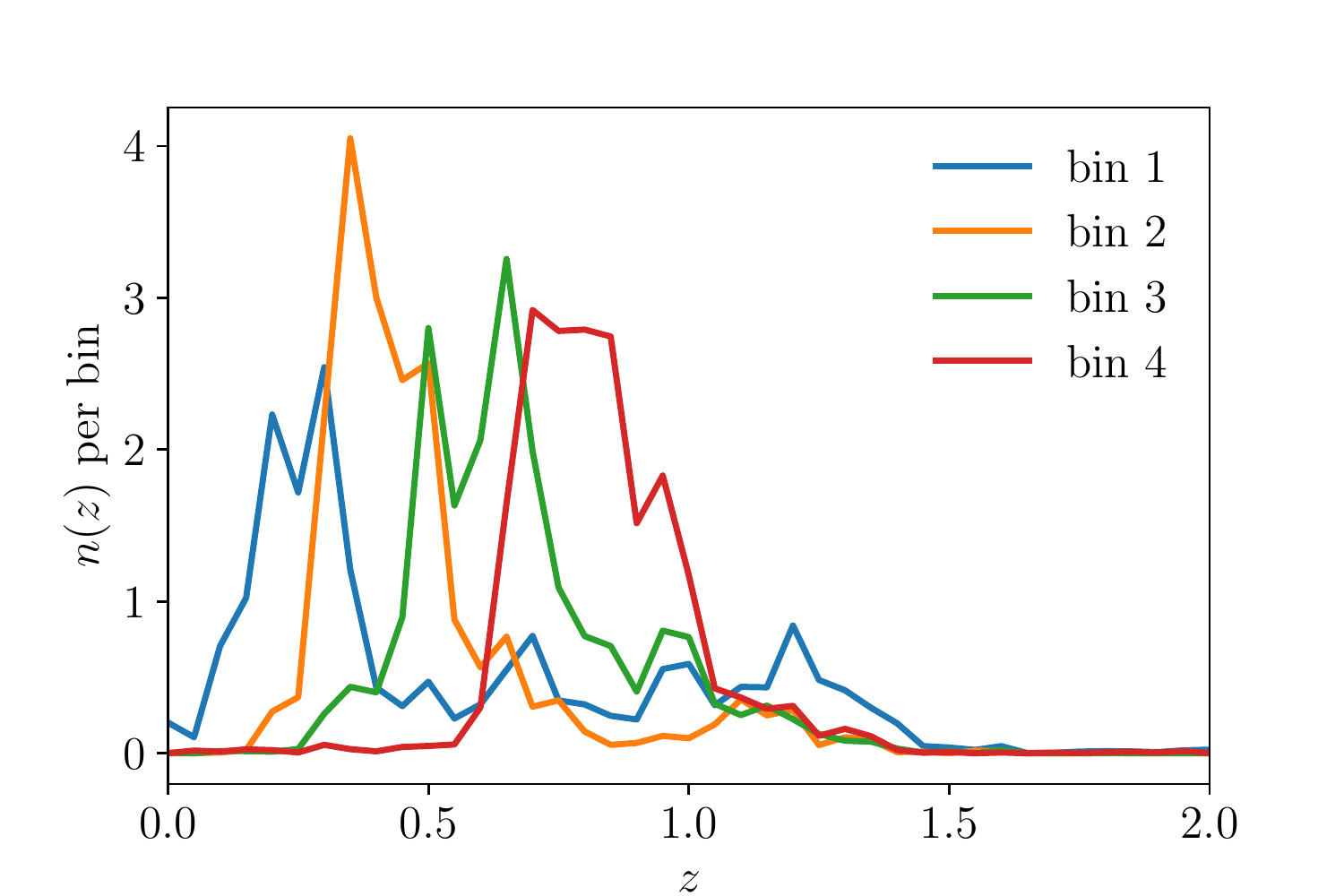}
\caption{Redshift distribution of sources in KiDS.}
\label{fig:redshift_distribution_KiDS}
\end{figure}

Now let us estimate quantitatively the finite-beam (or extended-source) corrections to the lensing power spectra and correlation functions, in a more realistic setting than Fig.~\ref{fig:damping_power_spectrum}. We generate the matter power spectrum~$P_\delta(\eta, k)$ with \textsc{camb}\footnote{\href{https://camb.info}{\tt https://camb.info}}, which integrates \textsc{halofit} for nonlinear scales. Cosmological parameters are the default \textsc{camb} parameters, corresponding to the \textsl{Planck} 2015 results~\cite{Ade:2015xua}. The redshift distribution of sources~$p(z)$ is chosen identical to the set of all four bins of the KiloDegree Survey (KiDS)~\footnote{\href{http://kids.strw.leidenuniv.nl/cosmicshear2016.php}{\tt http://kids.strw.leidenuniv.nl/cosmicshear2016.php}}, which is depicted in Fig.~\ref{fig:redshift_distribution_KiDS}.

Regarding the size of the sources, we assume that all galaxies are identical disks with physical radius~$R=10\U{kpc}$, but with random orientation. As a result, their projected area is $A = \pi R^2|\cos\iota|$, where $\iota$ is the inclination angle with respect to the line of sight. If orientation is random, then $\cos\iota$ is homogeneously distributed between $-1$ and $1$. Since we assumed in the present article that sources are circular, we account for this inclination effect by defining an effective radius~$r$, such that $\pi r^2 = A$. Thus, the probability distribution of this radius is linear,
\begin{equation}
p(r) = \frac{2 r}{R^2} \, [0\leq r \leq R] \ .
\end{equation}
We then use Eq.~\eqref{eq:joint_PDF} to determine $p(\beta, \chi_*)$.

\begin{figure}[h!]
\centering
\includegraphics[width=\columnwidth]{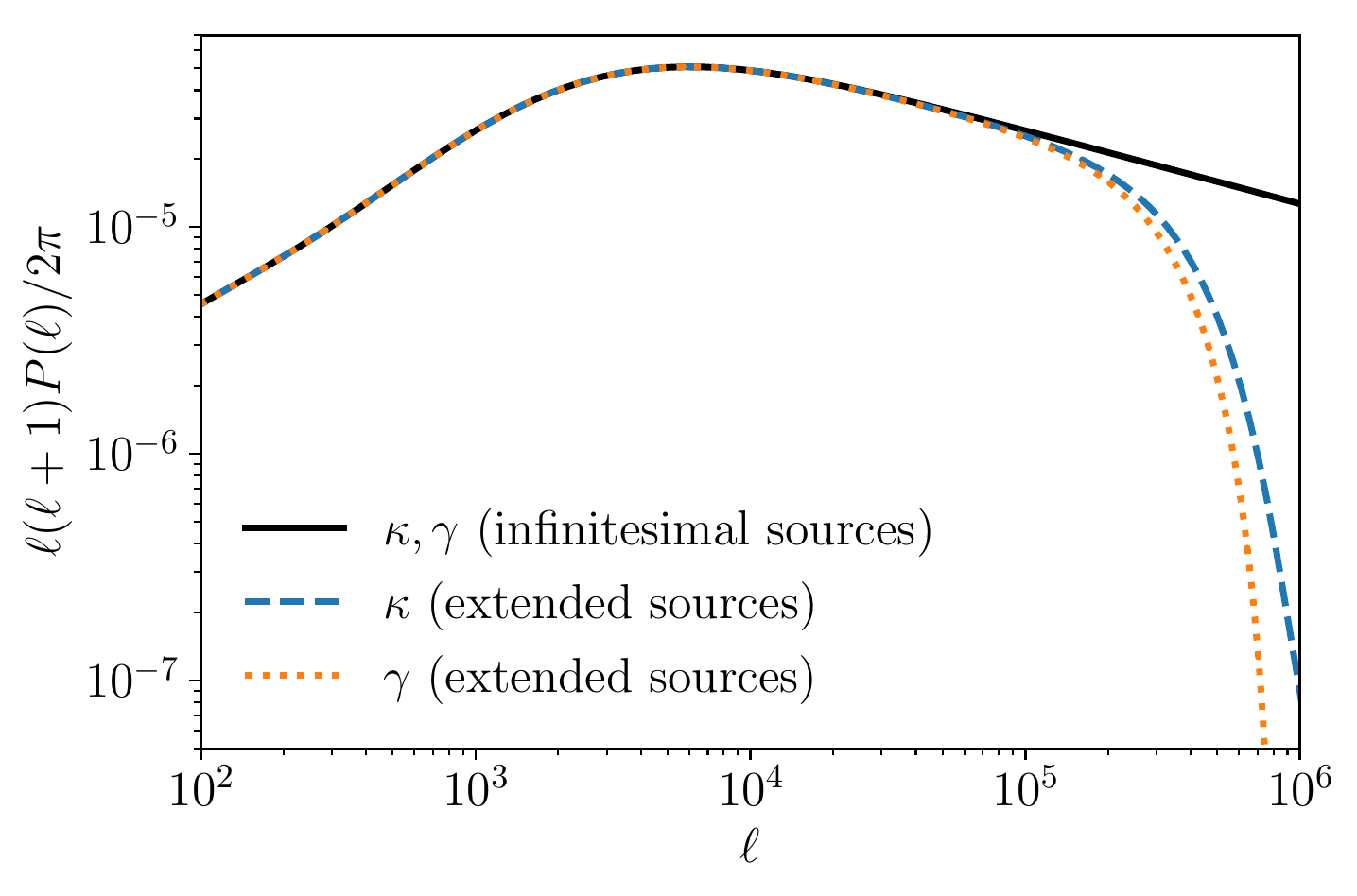}
\caption{Power spectra of cosmic convergence and cosmic shear as a function of multipole~$\ell$. The standard, infinitesimal-beam case, where the power spectrum of convergence and shear are equal, is shown by a solid black line. The blue dashed line and orange dotted line correspond, respectively, to convergence and shear when sources are extended. In these last two cases, sources have a physical size of $10\U{kpc}$, randomly oriented, and distributed in redshift as in the KiDS.}
\label{fig:power_spectra}
\end{figure}

These were the necessary ingredients for computing the integration kernels~$\bar{q}_\kappa$ and $\bar{q}_\gamma$ defined by Eqs.~\eqref{eq:kernel_convergence}, \eqref{eq:kernel_shear}, respectively, and the resulting convergence and shear power spectra for extended sources. Power spectra are shown in Fig.~\ref{fig:power_spectra}. As expected from the more qualitative discussions of Secs.~\ref{sec:convergence}, \ref{sec:shear}, the extension of sources tends to cut the power of both convergence and shear from $\ell>10^5$, which corresponds to angular scales of the order of $0.1\U{arcmin}$ and below. On those scales, the correlations of shear are more strongly reduced than the correlations of convergence.

\begin{figure}[h!]
\centering
\includegraphics[width=\columnwidth]{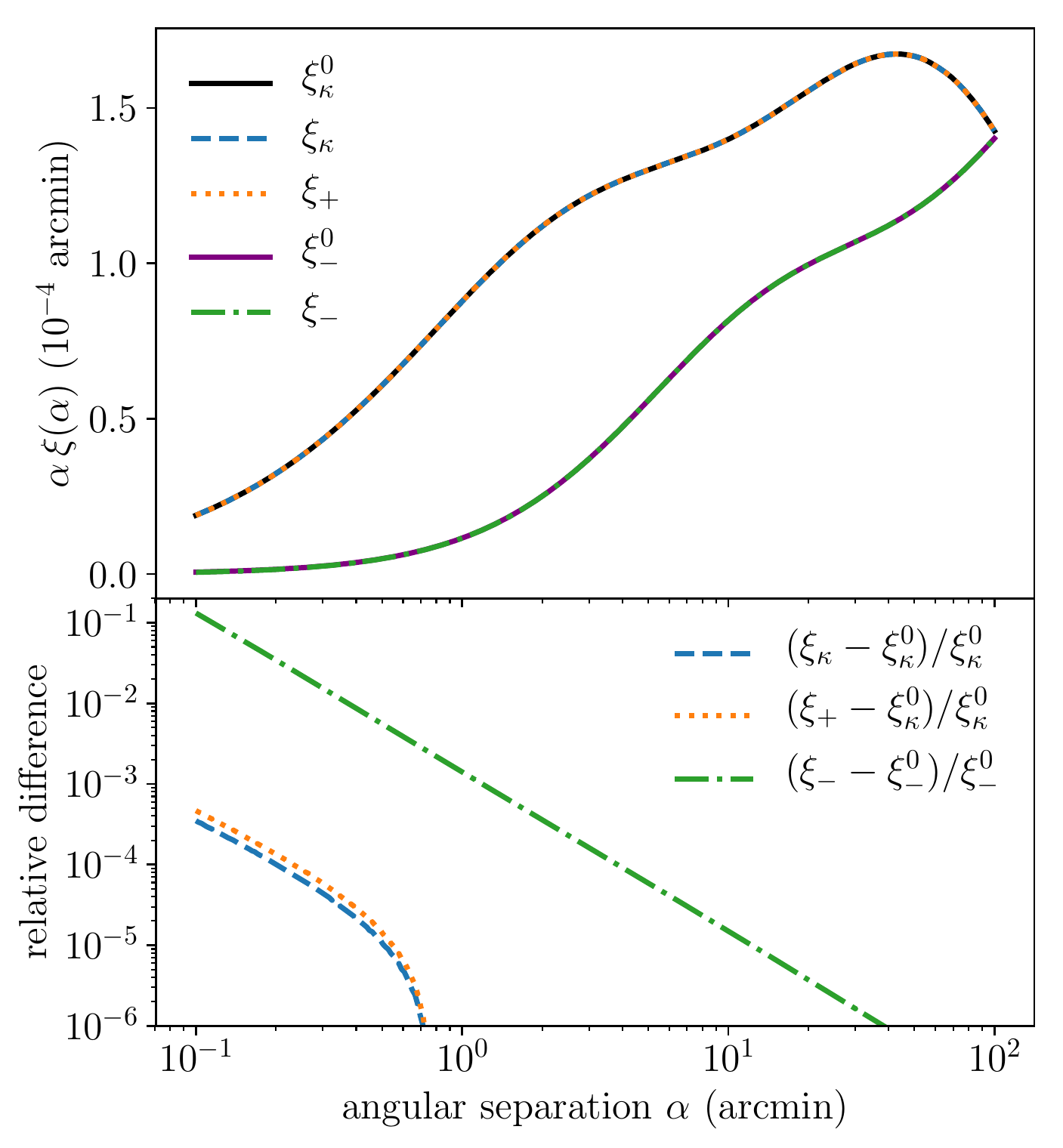}
\caption{Top panel: correlation functions of convergence ($\xi_\kappa$) and shear ($\xi_+$, $\xi_-$), defined in Secs.~\ref{sec:convergence}, \ref{sec:shear}, as functions of the angular separation~$\alpha$ of the two lines of sight. A $0$ superscript indicates the infinitesimal-source case, and the associated correlation functions are indicated by solid lines, black for $\xi_\kappa^0=\xi_+^0$ (upper curve), and purple for $\xi_-^0$ (lower curve). Correlation functions with extended sources, $\xi_\kappa, \xi_+, \xi_-$ are shown, respectively, with blue dashed, orange dotted, and green dot-dashed lines. Finite-source effects being invisible on the top, the bottom panel shows the relative corrections they induce.}
\label{fig:correlation_functions}
\end{figure}

Correlation functions are depicted in Fig.~\ref{fig:correlation_functions}. The top panel shows the correlation functions~$\xi_\kappa, \xi_+, \xi_-$ for both infinitesimal and extended sources. The lines being mostly superimposed, we also depict in the bottom panel the relative correction induced by finite-beam effects. These relative corrections remain smaller than $0.1\%$, for $\xi_\kappa, \xi_+$ and $\alpha\geq 0.1\U{arcmin}$; but they become of percent order and larger for $\xi_-$, as soon as $\alpha<0.5\U{arcmin}$. The fact that finite-beam corrections are more significant for $\xi_-$ is not surprising: this observable is known to be more sensitive to small scales than $\xi_+$. A percent-order systematic reduction of $\xi_-$ is, in principle, a considerable correction in the era of precision cosmology. However, one should keep in mind that such a reduction occurs at extremely small scales, where $\xi_-$ is already very small [$\xi_-(\alpha<1\U{arcmin})< 10^{-5}$], and which are not used for cosmological constraints in current lensing surveys. For example, KiDS~\cite{Hildebrandt:2016iqg} and DES~\cite{Abbott:2017wau} cut the range of angular separations~$\alpha$ below $4.2$ and $70\U{arcmin}$, respectively, when using $\xi_-$.

Better control of the small-scale physics and known systematics, like intrinsic alignments~\cite{Troxel:2014dba, Joachimi:2015mma}, could allow one to include smaller scales in future lensing analyses. However, even an ideal survey needs enough statistics to calculate correlations: the angular separation~$\alpha$ will always be bounded from below by the requirement that there are enough galaxies separated by $\alpha$. If we assume that the average galactic density today is $n_0\sim 0.5\U{Mpc^{-3}}$, then the projected surface density between $z=0$ and $z=2$ is~$n\e{proj}\approx 50\U{arcmin^{-2}}$, so that $\alpha\e{min} = 1/\sqrt{n\e{proj}} \approx 0.15\U{arcmin}$ is the absolute smallest angular scale ever reachable for a lensing survey with comparable depth. This implies that finite-beam corrections to $\xi_\kappa, \xi_+$ will most presumably remain below $0.1\%$, but percent-order corrections to $\xi_-$ cannot be excluded in futuristic surveys.

\section{Aperture masses}
\label{sec:aperture_masses}

Aperture mass is a lensing observable designed to break the so-called mass-sheet degeneracy (see e.g.~Ref.\cite{2001PhR...340..291B}). It is a useful tool to characterize the mass profile of galaxy clusters; in the context of weak lensing, it is more sensitive to small-scale correlations than~$\xi_+, \xi_-$, and thereby adapted to determining the small-scale behavior of the matter power spectrum. As a consequence, we expect aperture mass to be more affected by extended-source effects than other standard lensing observables.

\subsection{Definitions}

Consider an arbitrary line of sight, taken to coincide with the origin~$\vect{0}$ of angular positions~$\vect{\alpha}$. The aperture mass is defined as a weighted average of the convergence over a region of the sky with characteristic aperture~$\alpha\e{c}$,
\begin{equation}\label{eq:aperture_mass_kappa}
M\e{ap}^\kappa(\alpha\e{c})
\define \int \frac{\dd^2\vect{\alpha}}{\pi\alpha\e{c}^2} \; U\pa{\frac{\alpha}{\alpha\e{c}}} \, \kappa(\vect{\alpha}),
\end{equation}
where $U$ is a compensated filter,\footnote{We used a slightly different convention for $U$ compared to the literature (e.g. Refs.~\cite{2001PhR...340..291B, 2006glsw.conf.....M}), where the factor $\pi\alpha\e{c}^2$ at the denominator of Eq.~\eqref{eq:aperture_mass_kappa} is usually included in $U$. This propagates to the convention for $Q$ in Eq.~\eqref{eq:aperture_mass_gamma}.} that is,
\begin{equation}
\int \dd^2\vect{x} \; U(x) = 0.
\end{equation}
In Eq.~\eqref{eq:aperture_mass_kappa}, we added a superscript~$\kappa$ to the standard notation~$M\e{ap}$, because aperture mass can also be expressed in terms of tangential shear. Both expressions agree in the standard infinitesimal-beam case, but they do not for finite beams.

Let us elaborate on this. As a spin-two quantity, complex shear can be split into $+$ and $\times$ components. The $+$ component, usually called ``tangential'' and denoted~$\gamma\e{t}$, is in the present context defined as
\begin{equation}
\gamma\e{t}(\vect{\alpha}) \define -\Re\pac{\gamma(\vect{\alpha}) \ex{-2\i \omega}},
\end{equation}
where $\omega$ denotes the polar angle of $\vect{\alpha}=\alpha(\cos\omega, \sin\omega)$. For infinitesimal beams, the average of the tangential shear over an arbitrary circle turns out to be related to the average of convergence over the same circle and over the disk that it encloses. This property can be stated in a quite formal way as follows. Define the average of a function~$f(\vect{\alpha})$ over a circle~$\cir(\alpha)$ with radius $\alpha$ as
\begin{equation}
\ev{f}_{\cir(\alpha)} \define \int_0^{2\pi} \frac{\dd\omega}{2\pi} \; f(\vect{\alpha}) \ ,
\end{equation}
and the functional~$\mathcal{F}$ as, for any function $g$,
\begin{equation}
\mathcal{F}[g](\alpha) \define \frac{2}{\alpha^2} \int_0^{\alpha} \alpha' g(\alpha') \; \dd\alpha' - g(\alpha) \ ,
\end{equation}
then, for infinitesimal beams, we have the remarkable property~\cite{Bartelmann:1995yq}
\begin{equation}\label{eq:circular_averages}
\ev{\gamma\e{t}}_{\cir(\alpha)} = \mathcal{F}[\ev{\kappa}_{\cir(\alpha)}] \ .
\end{equation}

Equation~\eqref{eq:circular_averages} is, at first sight, a complicated way of writing the more common $\ev{\gamma\e{t}}_{\cir(\alpha)} = \ev{\kappa}_{\disk(\alpha)}- \ev{\kappa}_{\cir(\alpha)}$, where $\disk(\alpha)$ denotes the disk delimited by the circle $\cir(\alpha)$. Yet, it will ease the formulation of the rest of this section. Indeed, the functional~$\mathcal{F}$ is a sort of isometry: for any integrable function~$g$ and any compensated filter~$U$,
\begin{equation}\label{eq:isometry}
\int_0^\infty \alpha U(\alpha) g(\alpha) \; \dd\alpha
=
\int_0^\infty \alpha \mathcal{F}[U](\alpha) \mathcal{F}[g](\alpha) \; \dd\alpha \ ,
\end{equation}
as can be shown by integrations by parts. Substituting~$g(\alpha)=\ev{\kappa}_{\cir(\vect{\alpha})}$ in Eq.~\eqref{eq:isometry} and using Eq.~\eqref{eq:circular_averages}, we then obtain
\begin{equation}
M\e{ap}^\kappa(\alpha) = M\e{ap}^\gamma(\alpha) \quad \text{(infinitesimal sources)},
\end{equation}
where
\begin{equation}\label{eq:aperture_mass_gamma}
M\e{ap}^\gamma(\alpha\e{c}) \define \int \frac{\dd^2\vect{\alpha}}{\pi\alpha\e{c}^2} \; Q\pa{\frac{\alpha}{\alpha\e{c}}} \gamma\e{t}(\vect{\alpha}) \ ,
\quad
Q \define \mathcal{F}[U] \ .
\end{equation}
For extended sources, however, Eq.~\eqref{eq:circular_averages} is affected in a nontrivial way, so that $M\e{ap}^\kappa\not= M\e{ap}^\gamma$ in general.

\subsection{Aperture-mass standard deviation}

By virtue of the statistical homogeneity and isotropy of the Universe, the sky average of the aperture mass vanishes, but its standard deviation is a useful probe of the inhomogeneity of the matter distribution, especially on small scales. We here derive the extended-source corrections to this observable.

First of all, considering $M\e{ap}^\kappa$, it is straightforward to show that
\begin{equation}
\ev[2]{[M\e{ap}^\kappa(\alpha\e{c})]^2} = 2\pi \int_0^\infty \dd\ell \; \ell J^2(\ell\alpha\e{c}) P_\kappa(\ell) \ ,
\end{equation}
with
\begin{equation}\label{eq:def_J}
J(x) \define \frac{1}{\pi}\int_0^\infty \dd y \; y \, U(y) \, J_0(x y) \ ,
\end{equation}
regardless of whether the sources are infinitesimal or extended. Finite-beam corrections to $\ev[2]{[M\e{ap}^\kappa]^2}$ are thus fully encoded in the convergence power spectrum. Things go similarly for the shear, but the proof deserves further details. We start from the definition~\eqref{eq:aperture_mass_gamma}, and notice that the $\Re$ operator can be removed from the expression of $\gamma\e{t}$ thanks to the symmetry of the angular integration, so
\begin{multline}\label{eq:variance_M_ap_calculation}
\ev[2]{[M\e{ap}^\gamma(\alpha\e{c})]^2}
= \int \frac{\dd^2\vect{\alpha}_1}{\pi\alpha\e{c}^2} \frac{\dd^2\vect{\alpha}_2}{\pi\alpha\e{c}^2}
	 \; Q\pa{\frac{\alpha_1}{\alpha\e{c}}} Q\pa{\frac{\alpha_2}{\alpha\e{c}}} \\
	\times \ex{-2\i(\omega_1+\omega_2)}\ev{\gamma(\vect{\alpha}_1)\gamma(\vect{\alpha}_2)} \ .
\end{multline}
We then rewrite the correlation function using, e.g., Eqs.~\eqref{eq:xi_minus_def}, \eqref{eq:xi_minus_expr} and the definition of the Bessel function $J_4$. Calling $\Delta\vect{\alpha}\define \vect{\alpha}_1-\vect{\alpha}_2$, it goes as
\begin{align}
\ev{\gamma(\vect{\alpha}_1)\gamma(\vect{\alpha}_2)}
&= \ex{4\i\psi_{\Delta\vect{\alpha}}} \xi_-(\Delta\alpha) \\
&= \ex{4\i\psi_{\Delta\vect{\alpha}}} \int_0^\infty \frac{\dd\ell}{2\pi} \, \ell 
		J_4(\ell \Delta\alpha) P_\gamma(\ell) \\
&= \int\frac{\dd^2\vect{\ell}}{(2\pi)^2} \, \ex{\i\ell\cdot\Delta\vect{\alpha}+4\i\psi_{\vect{\ell}}} P_\gamma(\ell) \ ,
\end{align}
where $\psi_{\Delta\vect{\alpha}}, \psi_{\vect{\ell}}$ respectively denote the polar angles of the vectors~$\vect{\alpha}, \vect{\ell}$. Inserting the above expression into Eq.~\eqref{eq:variance_M_ap_calculation}, and using
\begin{multline}
\int\frac{\dd^2\vect{\alpha}}{\pi\alpha\e{c}^2} \; 
	Q\pa{\frac{\alpha}{\alpha\e{c}}} \ex{-2\i\omega\pm \i\vect{\ell}\cdot\vect{\alpha}} \\
= -2\pi \ex{-2\psi_{\vect{\ell}}} \int_0^\infty \frac{\alpha \dd\alpha}{\pi \alpha\e{c}^2} \, Q\pa{\frac{\alpha}{\alpha\e{c}}} J_2(\ell\alpha) \ ,
\end{multline}
we get
\begin{multline}
\ev[2]{[M\e{ap}^\gamma(\alpha\e{c})]^2} \\
= 2\pi \int_0^{2\pi} \dd\ell \; \ell 
	\pac{ \frac{1}{\pi}\int_0^\infty \dd y \; y \, Q(y) J_2(y \ell\alpha\e{c})}^2 P_\gamma(\ell) \ .
\end{multline}
The last step consists in noticing, from the property~\eqref{eq:isometry} of the functional $\mathcal{F}$, that the square bracket in the above expression is equal to $J(\ell\alpha\e{c})$, as defined in Eq.~\eqref{eq:def_J},
\begin{align}
J(x)
&= \frac{1}{\pi}\int_0^\infty \dd y \; y \, U(y) \, J_0(x y) \\
&= \frac{1}{\pi}\int_0^\infty \dd y \; y \, \mathcal{F}[U](y) \, \mathcal{F}[J_0](x y) \\
&= \frac{1}{\pi}\int_0^\infty \dd y \; y \, Q(y) \, J_2(x y) \ ;
\end{align}
whence
\begin{equation}
\ev[2]{[M\e{ap}^\gamma(\alpha\e{c})]^2} = 2\pi \int_0^\infty \dd\ell \; \ell J^2(\ell\alpha\e{c}) P_\gamma(\ell) \ .
\end{equation}
Therefore, just like $\ev[2]{[M\e{ap}^\kappa]^2}$, any extended-source correction to $\ev[2]{[M\e{ap}^\gamma]^2}$ is encoded in the power spectrum~$P_\gamma(\ell)$.

\begin{figure}[t]
\centering
\includegraphics[width=\columnwidth]{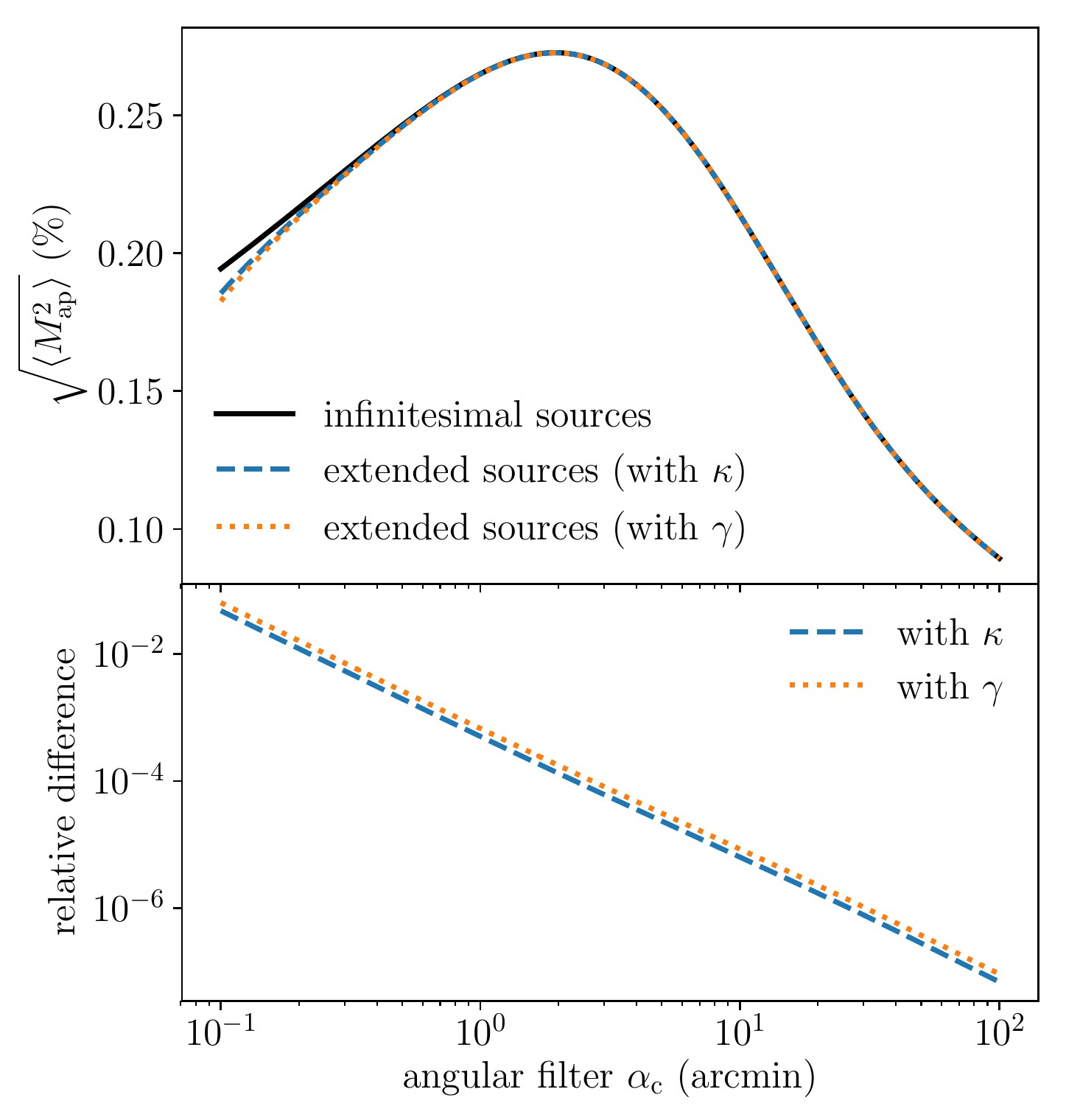}
\caption{Cosmic standard deviation of aperture mass~$M\e{ap}$, as a function of the aperture angle~$\alpha\e{c}$. In the top panel, a black line indicates the infinitesimal-source case, a blue dashed line corresponds to $M\e{ap}^\kappa$ for extended sources, and an orange dotted line shows $M\e{ap}^\gamma$ for extended sources as well. The bottom panel shows the relative difference between the infinitesimal-source and extended-source cases.}
\label{fig:aperture_mass}
\end{figure}

Figure~\ref{fig:aperture_mass} illustrates those corrections. The lensing power spectra are generated following the same procedure as in Sec.~\ref{sec:quantitative}, while the compensated filter~$U$ is chosen to be
\begin{equation}
U(x) \define 3\pa{1-x^2}\pa{ 1 - 3 x^2 } [0\leq x \leq 1] \ ,
\end{equation}
so that
\begin{equation}
J(x) = \frac{12}{\pi} \frac{J_4(x)}{x^2} \ .
\end{equation}
Similarly to $\xi_+, \xi_-$, we see that extended-source corrections to the aperture mass tend to manifest on subarcmin scales; they reach percent level for $\alpha\e{c}\leq 0.2\U{arcmin}$.

It is apparent from the bottom panel of Fig.~\ref{fig:aperture_mass} that the relative corrections to $M\e{ap}^\kappa, M\e{ap}^\gamma$ are proportional. More precisely,
\begin{equation}\label{eq:the_other_4/3}
\frac{\ev[2]{[M\e{ap}^\gamma]^2} - \ev[2]{[M\e{ap}^0]^2}}
		{\ev[2]{[M\e{ap}^\kappa]^2} - \ev[2]{[M\e{ap}^0]^2}} \approx \frac{4}{3} \ ,
\end{equation}
where, again, a $0$ superscript denotes the infinitesimal-beam case. This $4/3$ is \emph{not} the same as the one discussed in Sec.~\ref{sec:4/3}. Instead, it is related to the behavior of the finite-beam corrections to the lensing power spectra for $\ell\beta\ll 1$. This can be understood as follows: consider the difference
\begin{multline}
\ev[2]{[M\e{ap}^{\kappa}]^2} - \ev[2]{[M\e{ap}^0]^2}  \\
= 2\pi \int_0^\infty \dd\ell \; \ell J^2(\ell\alpha\e{c}) \pac{P_{\kappa}(\ell) - P_\kappa^0(\ell)} \ .
\end{multline}
The aperture-mass kernel~$J^2(\ell\alpha\e{c})$ peaks around $\ell\approx 4/\alpha\e{c}$. For angular apertures~$\alpha\e{c}$ much larger than the typical beam size~$\beta$, $J^2$ thereby selects multipoles such that~$\ell\beta\ll 1$. We have seen at the end of Sec.~\ref{sec:shear} that, for sources of equal apparent size~$\beta$, if $\ell\beta\ll 1$ then
\begin{equation}
P_\kappa(\ell) - P_\kappa^0(\ell) \approx \frac{(\ell\beta)^2}{4} \, P_\kappa^0(\ell) \ .
\end{equation}
Since the standard convergence power spectrum is essentially constant with the peak of $J^2$, we can thus write
\begin{multline}\label{eq:correction_M_ap_kappa}
\ev[2]{[M\e{ap}^\kappa]^2} - \ev[2]{[M\e{ap}^0]^2} \\
\approx \frac{1}{4} \times 2\pi P_\kappa^0(4/\alpha\e{c}) \int_0^\infty \dd\ell \; \ell J^2(\ell\alpha\e{c}) (\beta\ell)^2 \ .
\end{multline}
A similar reasoning applies to $M\e{ap}^\gamma$. In this case, since
\begin{equation}
P_\gamma(\ell) - P_\kappa^0(\ell) \approx \frac{(\ell\beta)^2}{3} \, P_\kappa^0(\ell) \ ,
\end{equation}
the prefactor $1/4$ in Eq.~\eqref{eq:correction_M_ap_kappa} must be replaced by $1/3$, whence the $4/3$ of Eq.~\eqref{eq:the_other_4/3}.

\section{Conclusion}
\label{sec:conclusion}

In this article, we evaluated the finite-beam (extended-source) corrections to the standard weak-lensing observables, namely the convergence and shear two-point correlations. We found that these correlations are damped on very small scales, i.e. when the angular separation of the sources becomes comparable to the size of the sources themselves. As a rule of thumb, the angular power spectra of convergence and shear read
\begin{align}
P_\kappa(\ell) &\approx \pac{\frac{2J_1(\ell\beta)}{\ell\beta}}^2 P_\kappa^0(\ell) \ ,\\
P_\gamma(\ell) &\approx \pac{\frac{4J'_2(\ell\beta)}{\ell\beta}}^2 P_\gamma^0(\ell) \ ,
\end{align}
where $J_n$ are Bessel functions, and $\beta$ is the typical angular radius of the sources, while $P_\kappa^0=P_\gamma^0$ correspond to the infinitesimal-source case.

If the Universe were made of discrete objects with Poisson distribution, then such corrections would imply the large effects found in \FLUletter, such as the $4/3$-violation of the Kaiser-Squires relation between shear and convergence. However, in reality, the matter distribution exhibits correlations on a wide range of scales, and only a tiny amount of these correlation turns out to be damped by finite-beam effects.

For sources with a redshift distribution comparable to the one of the KiDS, with an intrinsic size of $10\U{kpc}$, and random inclination, we found that significant corrections appear in the power spectrum from $\ell>10^5$. As a consequence, in the two-point correlation functions, corrections to both $\xi_\kappa$ and $\xi_+$ remain below $0.1~\%$ for angular separation larger than $0.1\U{arcmin}$; they can reach $1\%$ for $\xi_-$ but only on subarcmin scales. Corrections to the cosmic variance of the aperture mass are similar to $\xi_-$.

Therefore, finite-source corrections to the standard weak-lensing observables do not need to be considered in current and near-future surveys. They could be a significant systematic effect only if we could reach subarcmin scales, and if the other important known sources of systematics, such as intrinsic alignments, were under control on those scales. Our companion article \FLUadvanced~\cite{advanced} explores further aspects of the weak lensing of extended sources, where finite-size effects can play a more significant role.

\section*{Acknowledgements} It is a pleasure to thank Marika Asgari, Camille Bonvin, Alex Hall, Fabien Lacasa, Anthony Lewis, Ermis Mitsou, John Peacock, Cyril Pitrou, Nick Kaiser, Peter Schneider, and Elena Sellentin for stimulating discussions on this project. P.~F. acknowledges support by the Swiss National Science Foundation. The work of J.-P.~U. is made in the ILP LABEX (under Reference No. ANR-10-LABX-63) and was supported by French state funds managed by the ANR within the Investissements d'Avenir program under Reference No ANR-11-IDEX-0004-02. The work of J.~L. work is supported by the National Research Foundation (South Africa).

\appendix
\section{Lens equation and the choice of a background}
\label{app:negative_lenses}

This Appendix aims at justifying our, perhaps surprising, choice of allowing lenses to have negative masses in Sec.~\ref{sec:finite-beam_formalism}. This requires to go back to the derivation of the lens equation~\eqref{eq:lens_equation}. The first step, which is somehow the most important, consists in choosing a \emph{background}, that is, in the present case, a reference spacetime and a coordinate system such that light propagates in straight lines. This corresponds to a no-lensing situation, on the top of which  perturbers (the lenses) are added.

\subsection{One spacetime, two backgrounds}

Suppose one wants to describe a Universe filled with discrete masses. Then the most natural background is Minkowski, on the top of which one adds Newtonian perturbations,
\begin{equation}\label{eq:Minkowski_background}
\dd s^2 = -(1+2\Phi\e{N})\dd t^2 + (1-2\Phi\e{N}) \delta_{ij} \dd X^i \dd X^j \ ,
\end{equation}
where $\Phi\e{N}$ is the Newtonian potential; we assumed that the cosmological constant is zero for simplicity. If the masses are noncompact and slowly moving, then
\begin{equation}
\delta_{ij} \frac{\partial^2\Phi}{\partial X^i \partial X^j} = 4\pi G \rho \ .
\end{equation}
where $\rho$ is the actual matter density. In particular, for a discrete Universe, we have the usual Newtonian result
\begin{equation}\label{eq:Newtonian_potential}
\Phi\e{N}(t,\vect{X}) = - \sum_k \frac{G m_k}{|\vect{X}-\vect{X}_k(t)|} \ .
\end{equation}
The expansion of the Universe would then manifest as a Hubble radial motion of the masses, $\dd\vect{X}_k/\dd t=H(t) \vect{X}_k(t)$.

Although the Minkowski background is naturally called by the situation that we wish to model here---we used it in \FLUletter, it is known to suffer from important drawbacks when dealing with cosmic scales. The main one is probably that the Newtonian condition that objects are slowly moving (compared to the speed of light) is not satisfied at distances comparable to the Hubble radius~$H^{-1}$. Hence Eq.~\eqref{eq:Minkowski_background} cannot be safely applied beyond small regions in the Universe. Instead, one is then forced to consider the alternative FLRW background, which, supplemented with scalar perturbations, has a line element
\begin{equation}\label{eq:FLRW_background}
\dd s^2 = a^2(\eta)\pac{-(1+2\Phi)\dd \eta^2 + (1-2\Phi) \delta_{ij} \dd x^i \dd x^j },
\end{equation}
where we restricted to $K=0$ for an easier comparison. We used a different notation for the gravitational potential, because, contrary to $\Phi\e{N}$, $\Phi$ satisfies the shifted Poisson equation
\begin{equation}
\delta_{ij} \frac{\partial^2\Phi}{\partial x^i \partial x^j} = 4\pi G a^2 (\rho-\bar{\rho}) \ .
\end{equation}
The presence of $a^2$ in this equation only accounts for the difference between comoving coordinates ($x^i$) and the more physical coordinates~$X^i=a x^i$. Therefore, the main difference between $\Phi$ and $\Phi\e{N}$ is the Newtonian potential~$\bar{\Phi}$ created by the mean density of the Universe
\begin{equation}
\Phi = \Phi\e{N} - \bar{\Phi}
\end{equation}
with $\delta^{ij}\partial^2\bar{\Phi}/\partial X^i\partial X^j = 4\pi G \bar{\rho}$, so that $\bar{\Phi}\propto \bar{\rho} R^2$

Before moving to the lens equation, let us stress that both \emph{perturbed} geometries~\eqref{eq:Minkowski_background}, \eqref{eq:FLRW_background} describe the same spacetime, at least locally. However, since their respective \emph{background} ($\Phi\e{N}=0$ and $\Phi=0$) differs, what one would call ``lensing'' differs too, because lensing is a deflection of actual light rays with respect to the background.

\subsection{Lens equation and negative masses}

The null geodesic equation for the geometry~\eqref{eq:FLRW_background} can be analytically solved at first order in $\Phi$ (see, e.g., Sec.~5.2.1 of Ref.~\cite{Fleury}), and the result is
\begin{equation}
\vect{\theta} - \vect{\beta} = -2\int_0^{\chi\e{S}} \dd\chi \; \frac{\chi (\chi\e{S}-\chi)}{a(\chi)\chi\e{S}} 
																\, \vect{\nabla}_\perp \Phi \ ,
\end{equation}
where $\vect{\beta}$ is the unlensed position of the source in FLRW, $\chi\e{S}$ is the comoving radial coordinate of the source, and transverse gradient is based on physical coordinates, $\nabla^i \define \partial/\partial X^i$. The following step would then be simpler if the integrand only depended on $\Phi\e{N}$, which is well localized on the lenses, but as discussed above, $\Phi=\Phi\e{N}-\bar{\Phi}$, and hence the actual lens equation reads
\begin{multline}\label{eq:lens_equation_intermediate}
\vect{\theta} - \vect{\beta} = \sum_{k} \frac{4 G m_k (\chi\e{S} - \chi_k)}{\chi_k \chi\e{S}}
															\, \frac{\vect{\theta}-\vect{\lambda}_k}{|\vect{\theta}-\vect{\lambda}_k|^2}
															+ \Delta\vect{\beta} \ ,
\end{multline}
where $\Delta\vect{\beta}$ is the difference of the unlensed position in FLRW and Minkowski, due to the homogeneous density~$\bar{\rho}$ in FLRW,
\begin{equation}
\Delta\vect{\beta} = 2\int_0^{\chi\e{S}} \dd\chi \; \frac{\chi (\chi\e{S}-\chi)}{a(\chi)\chi\e{S}} 
																\, \vect{\nabla}_\perp \bar{\Phi} \ .
\end{equation}
More precisely, because $\Delta\vect{\beta}$ is generated by $-\bar{\Phi}$, it is the deflection introduced by a homogeneous \emph{negative} density~$-\bar{\rho}$. This term can thus be put under the same form as the first, regular, term of Eq.~\eqref{eq:lens_equation_intermediate}, by modeling $\bar{\rho}$ as a set of many small negative masses, homogeneously distributed in the Universe.

One could wonder why we tried so hard to keep an FLRW background, while using a discrete form for the lens equation. The advantage of the FLRW background is that it makes things much easier when introducing cosmological quantities like the matter power spectrum; it also makes our results more readable to the cosmology community. As for the lens equation, although we eventually take the continuous limit in Sec.~\ref{sec:discrete_to_continuous}, the discrete approach is pedagogically very superior. In particular, it makes the discussion about the respective roles of interior and exterior lenses clearer, and eases comparisons with \FLUletter.



\bibliography{bibliography_finite_beams}

\end{document}